\documentclass[fleqn,usenatbib]{mnras}

\usepackage{newtxtext,newtxmath}

\usepackage[T1]{fontenc}

\DeclareRobustCommand{\VAN}[3]{#2}
\let\VANthebibliography\thebibliography
\def\thebibliography{\DeclareRobustCommand{\VAN}[3]{##3}\VANthebibliography}

\usepackage{graphicx}	
\usepackage{amsmath}	
\usepackage{multicol}        
\usepackage{bm}		
\usepackage{pdflscape}	

\usepackage{multirow}
\usepackage{url}
\usepackage{array}
\usepackage{longtable}
\usepackage{tablefootnote}
\usepackage{subfigure}

\def\vlsr  {\ifmmode {V_{\rm LSR}}\else {$V_{\rm LSR}$}\fi}

\title[The Milky Way atlas for linear filaments]{
	The Milky Way atlas for linear filaments II. clump rotation versus filament orientation
	}

\author[Xuefang Xu et al.]{Xuefang Xu$^{1,2}$, Ke Wang$^{3}$\thanks{E-mail:kwang.astro@pku.edu.cn}, 
	Qian Gou$^{1,2}$\thanks{E-mail:qian.gou@cqu.edu.cn},Tapas Baug$^{4}$, 
	Di Li$^{5,6}$,  Chunguo Duan$^{1}$, and Juncheng Lei$^{1}$
\\
$^{1}$School of Chemistry and Chemical Engineering, Chongqing University, 
Chongqing 401331, China\\
$^{2}$Chongqing Key Laboratory of Chemical Theory and Mechanism, 
Chongqing University, Chongqing 401331, China\\
$^{3}$Kavli Institute for Astronomy and Astrophysics, Peking University, 
5 Yiheyuan Road, Haidian District, Beijing 100871, China\\
$^{4}$S. N. Bose National Centre for Basic Sciences, Block JD, Sector III, 
Salt Lake, Kolkata 700106, India\\
$^{5}$Department of Astronomy, Tsinghua University, Beijing 100084, China\\
$^{6}$National Astronomical Observatories, Chinese Academy of Sciences, 
Beijing 100012, China\\
}

\begin{document}
\label{firstpage}
\pagerange{\pageref{firstpage}--\pageref{lastpage}}
\maketitle

\begin{abstract}
Dense clumps distributed along filaments are the immediate medium for star formation. Kinematic properties of the clumps, such as velocity gradient and angular momentum, combined with filament orientation, provide important clues to the formation mechanism of filament-clump configurations and the role of filaments in star formation. By cross-matching the Milky Way atlas for linear filaments and the Structure, Excitation and Dynamics of the Inner Galactic Interstellar Medium (SEDIGISM) $^{13}$CO (2-1) data, we aim to derive the velocity gradient (${\cal{}G}$) and its direction ($\theta_{\cal{}G}$), the specific angular momentum ($J/M$), and the ratio ($\beta$) between the rotational energy and gravitational energy of clumps, as well as to investigate the alignment between clump rotation and filament orientation. We found a monotonic increase in $J/M$ as a function of clump size ($R$), following a power-law relation $J/M~\propto~R^{1.5\pm0.2}$. The ratio $\beta$ ranges from 1.1~$\times$~10$^{-5}$ to 0.1, with a median value 1.0~$\times$~10$^{-3}$, suggesting that clump rotation provides insignificant support against gravitational collapse. The distribution of the angle between clump rotation ($\theta_{\cal{}G}$) and natal filament orientation is random, indicating that the clumps’ rotational axes have no discernible correlation with the orientation of their hosting filaments. Counting only the most massive clump in each filament also finds no alignment between clump rotation and filament orientation. 
\end{abstract}




\section{Introduction}

Filamentary structures are ubiquitous in the interstellar medium (ISM)~\citep{Molinari:10,Andre:14}. They are believed to play an important role in star formation, because dense molecular clumps and cores (the immediate medium for star formation) are predominantly observed along the filamentary ridges compared to other parts within a molecular cloud. Clumps along filaments often form a Necklace-like geometric configuration, which is a natural consequence of the so called ``sausage'' instability in a plasma/magnetized gaseous cylinder~\citep{Chandrasekhar:53}. ~\citet{Wang:16} developed a customized minimum spaning tree (MST) algorithm to identify such filaments by connecting clumps in position-position velocity (PPV) space into a velocity-coherent linear structure. They have applied the MST method to Galactic surveys (BGPS, ALASGAL, SEDIGISM, and HiGAL) and revealed a panoramic view of the filamentary structures in the Galaxy~\citep{Wang:16,Ge:22,Ge:23,Wang:24}. Of particular interest to this study is the first Milky Way atlas of 42 linear filaments in the full Galactic plane~\citep{Wang:24}, providing well-defined filament orientation in an unbiased sample. In the Necklace configuration, filaments can channel gas flows to feed star formation in the clumps, which has been observed in many filaments~\citep[e.g.][]{Wang:14,Ge:22,Ge:23,Xu:23}. 

In sufficiently young filaments, such as those identified in \textit{Herschel} far-IR emission of cold dust~\citep{Lid:12,Wang:15,Wang:24}, it is believed that the clumps may still retain the kinematic signatures inherited from  the cylindrical fragmentation~\citep{Shimajiri:19,Sharma:20}.  For example,~\cite{Ren:23} reported that the central massive core  in massive filament G352.63-1.07 has a spatial extent parallel to  the main filament. On the other hand,  simulations~\citep{Misugi:19,Misugi:23} found that  most of the cores, formed from filament fragmentation,  rotate perpendicular to the filament's long-axis.  This may be due to the fact that gas in filaments predominantly  flows either perpendicular or parallel to the filament axis  as also observed by~\citet{Sharma:20}. Classically, the angular momentum of a star-forming molecular cloud is anticipated to be hierarchically transferred to the clumps/cores and  eventually to the protostar. As the gas flow is believed to occur either parallel or perpendicular to the long-axis of the filament, the rotation axis (of the clumps, core or accretion disk) is expected to be preferentially parallel or perpendicular to the core minor axis and the filament structure. Thus, the orientation of clump angular momentum and their host filaments may provide important clues about their formation mechanisms and the role of filaments in star formation. 

To explore the alignment between clump rotation and the natal environment, the distribution of angles between the rotational axes of clump and the orientation of natal filament were analyzed. Such angles could potentially differentiate between various mechanisms of clump formation. For instance, the angular momentum of a clump is oriented perpendicularly to its natal filament’s axis~\citep[e.g.][]{Anathpindika:08,Wang:11,Kong:19}, providing evidence for the clump-formation mechanism proposed by~\citet{Whitworth:95}. Conversely, a near parallel relation can be clarified by another clump formation mechanism, namely Gravo-turbulent fragmentation~\citep{Banerjee:06,Offner:08,Clarke:17,Aizawa:20,Anathpindika:22}. Furthermore,~\citet{Baug:20} found a random distribution between angular momentum and filament orientation. In these cases, the direction of angular momentum is traced by outflow or disk axes. Velocity gradients were fitted by~\citet{Xu:20a} following~\citet{Goodman:93} to trace the rotational direction of cores, revealing no preferred trend between cores and their natal filaments specially in the Orion Molecular Cloud (OMC)-2 and OMC-3 filamentary regions~\citep{Lid:03,Lid:13,Yue:21}. Additionally, their defined orientation of gas-flow axis, as per the auto-correlation function in the N$_{2}$H$^{+}$ integrated density map~\citep[similar to][]{Li:13}, did not show any preferred relation with the embedded cores.  

In this work, the alignment between clump rotation and the natal filament in eight linear filaments was investigated. The paper is organized as follows: Section~\ref{sec:da} outlines the target filaments selected from an unbiased linear filament sample~\citep{Wang:24} and describes the $^{13}$CO (2-1) data used in this study. Section~\ref{sec:fila} details the definitions of the orientations of the target filaments, while Section~\ref{sec:gradient} discusses the velocity gradient measurements of HiGAL clumps~\citep{Elia:21} following the methodology established by~\citet{Goodman:93}. Section~\ref{sec:res} shows the result of the specific angular momentum ($J/M$), the ratio ($\beta$) between the rotational energy and gravitational energy, and the angles between filaments and clump velocity gradients ($\vert{}\theta_{f} - \theta_{\cal{}G}{}\vert$). In Section~\ref{sec:dis}, we compare $J/M$ and $\beta$ in this work with that in previous works, and discuss the distribution of $\vert{}\theta_{f} - \theta_{\cal{}G}{}\vert$. The main results are summarized in Section~\ref{sec:sum}. 

\section{Data and Methodology}\label{sec:dm}
\subsection{Filament sample and Data} \label{sec:da}
The unbiased linear filament sample, the first large-scale linear filaments across the full Galactic plane~\citep{Wang:24}, was combined with the SEDIGISM survey (Structure, Excitation and Dynamics of the Inner Galactic Interstellar Medium), resulting eight target filaments (F2, F3, F34, F35, F36, F39, F40, and F41). The filament sample includes filaments of varying distances, lengths, masses, luminosities, and linearities~\citep{Wang:24}. 

Data from the SEDIGISM survey, conducted with the Atacama Pathfinder Experiment 12 m submillimetre telescope~\citep[APEX,][]{Gusten:06} were used. The survey overview papers~\citep{Schuller:17,Schuller:21,Duarte:21} provide a comprehensive account of the observations, data reduction, and data-quality checks. In total, the SEDIGISM survey observed 84 deg$^{2}$, covering from -60$^{\circ}{}\leq{}l{}\leq{}$+18$^{\circ}$, and $\vert{}b{}\vert{}\leq{}$0.5$^{\circ}$, with a few extensions in $b$ towards some regions, as well as an additional field towards the W43 region (+29$^{\circ}{}\leq{}l{}\leq{}$+31$^{\circ}$). The $^{13}$CO (2-1) images used here are from the DR1 data set~\citep{Schuller:21}, which has a typical 1$\sigma$ sensitivity of 0.8 $-$ 1.0\,K (in T$_{mb}$) per 0.25\,km\,s$^{-1}$ channel and an FWHM beam size of 28$''$.  

\subsection{Orientations of the Target Filaments} \label{sec:fila}
\begin{figure*}
	\centering
	\subfigure{
		\centering
		\includegraphics[width=0.55\textwidth]{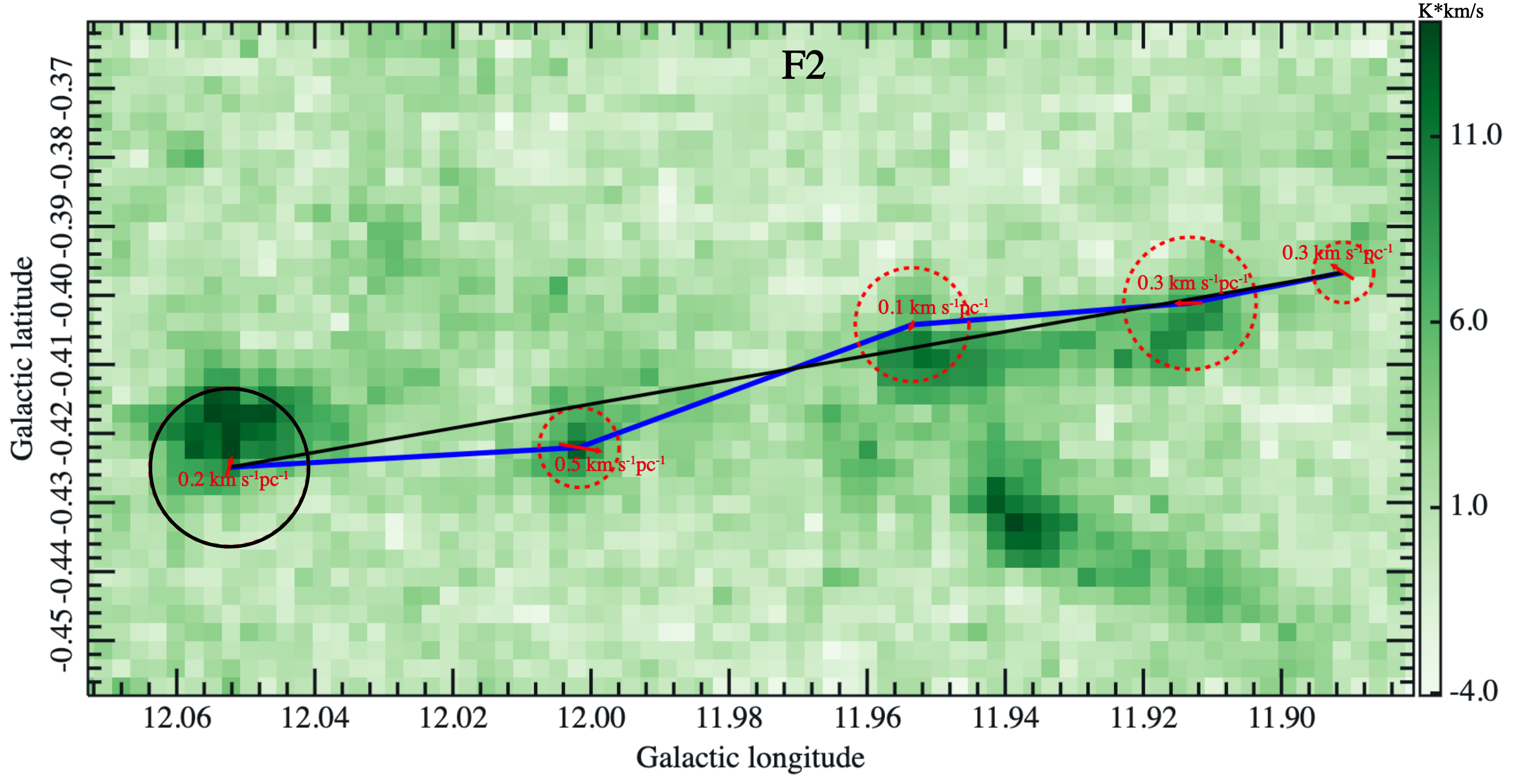}
		\includegraphics[width=0.42\textwidth]{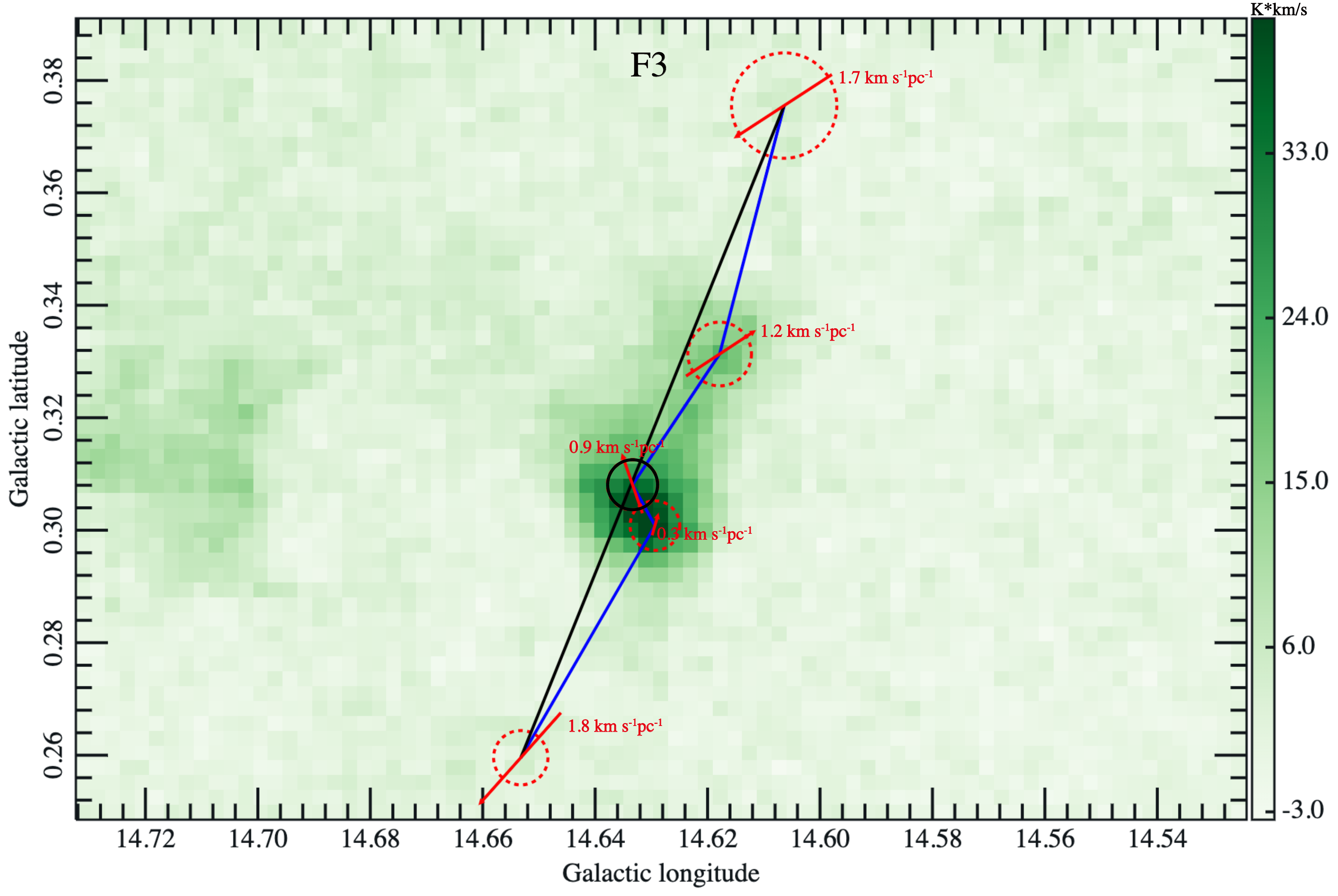}
	}%
	
	\subfigure{
		\centering
		\includegraphics[width=0.43\textwidth]{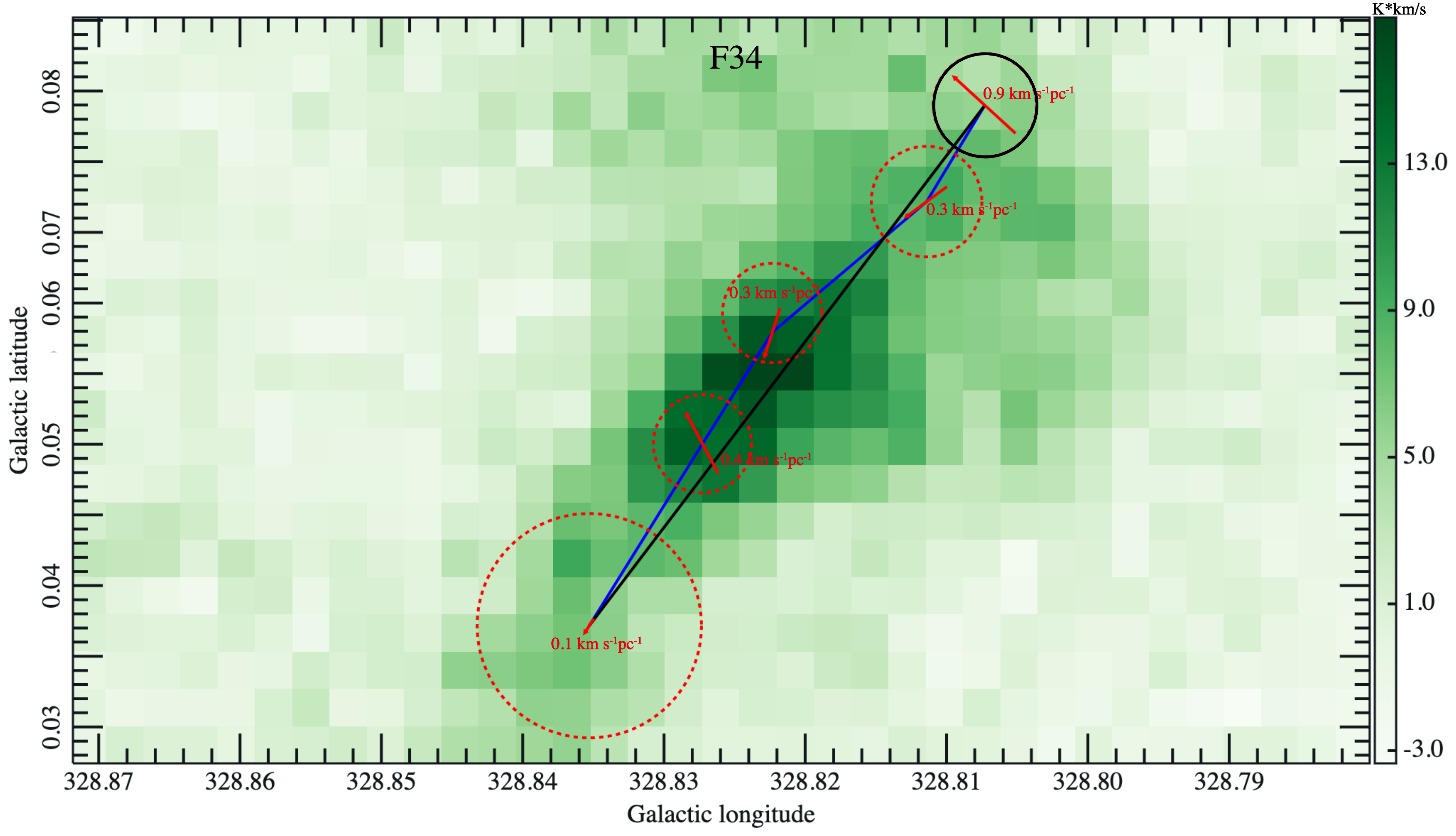}
		\includegraphics[width=0.53\textwidth]{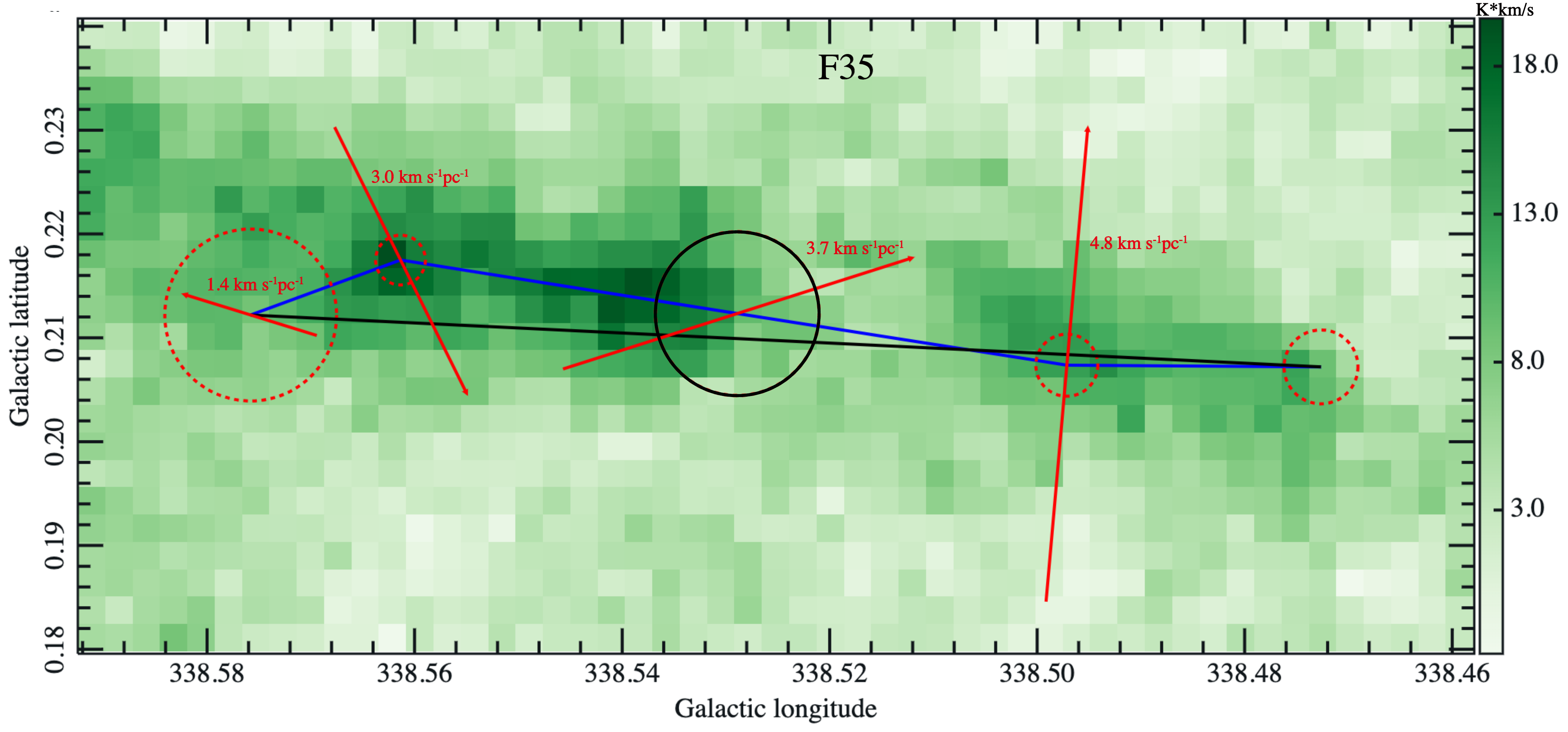}
	}%

	\subfigure{
		\centering
		\includegraphics[width=0.485\textwidth]{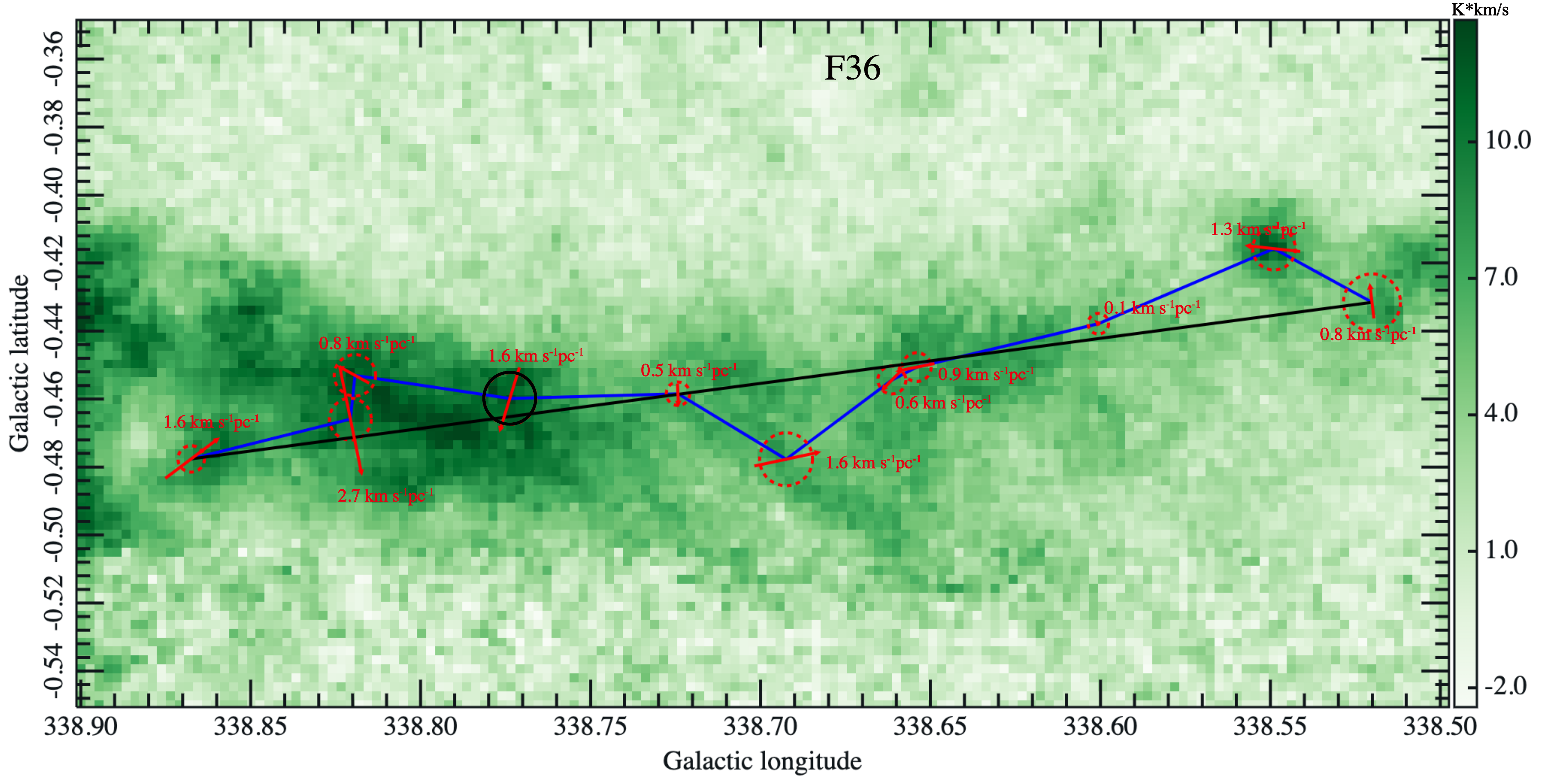}
		\includegraphics[width=0.49\textwidth]{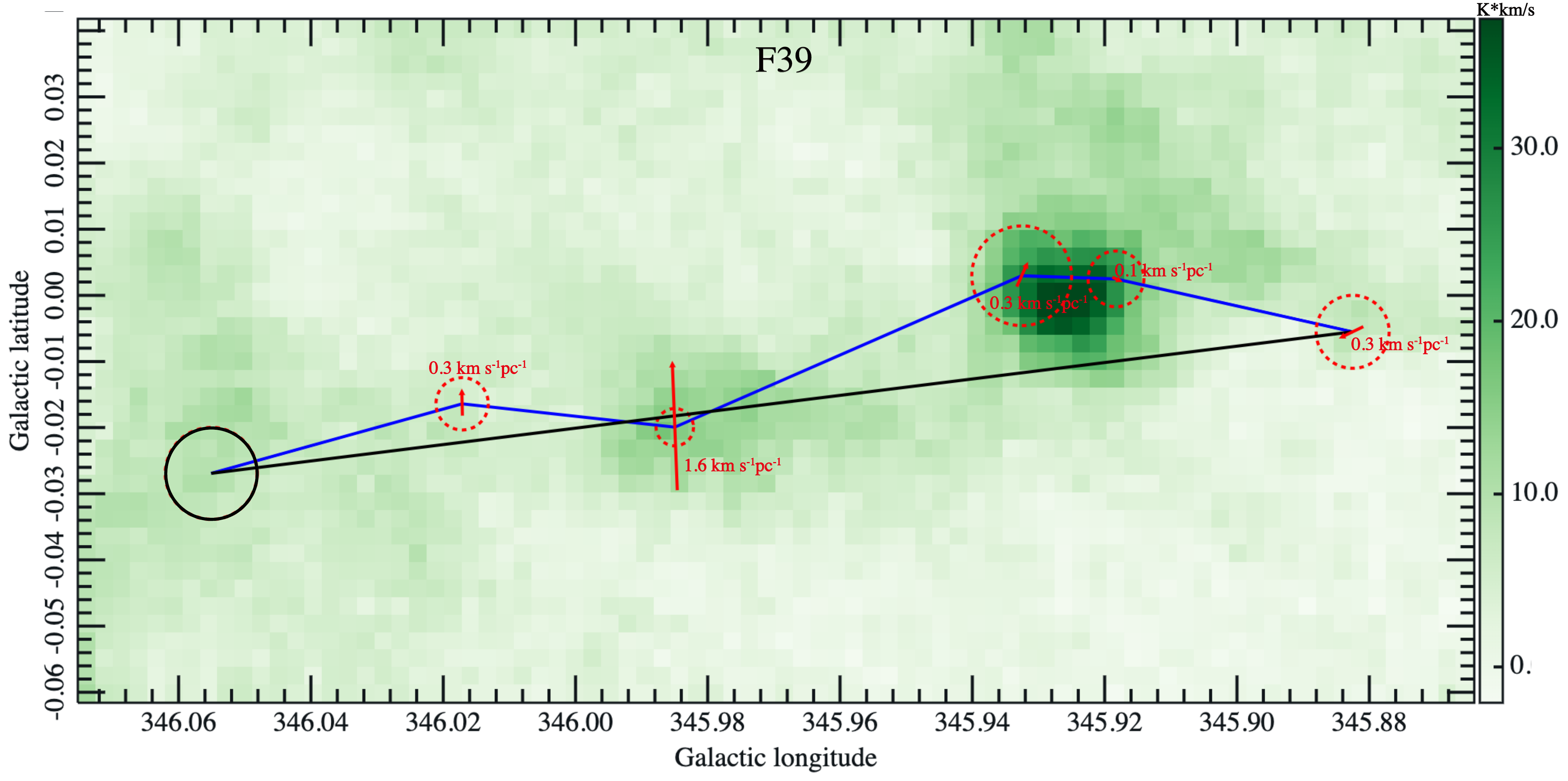}
	}%
	
	\subfigure{
		\centering
		\includegraphics[width=0.54\textwidth]{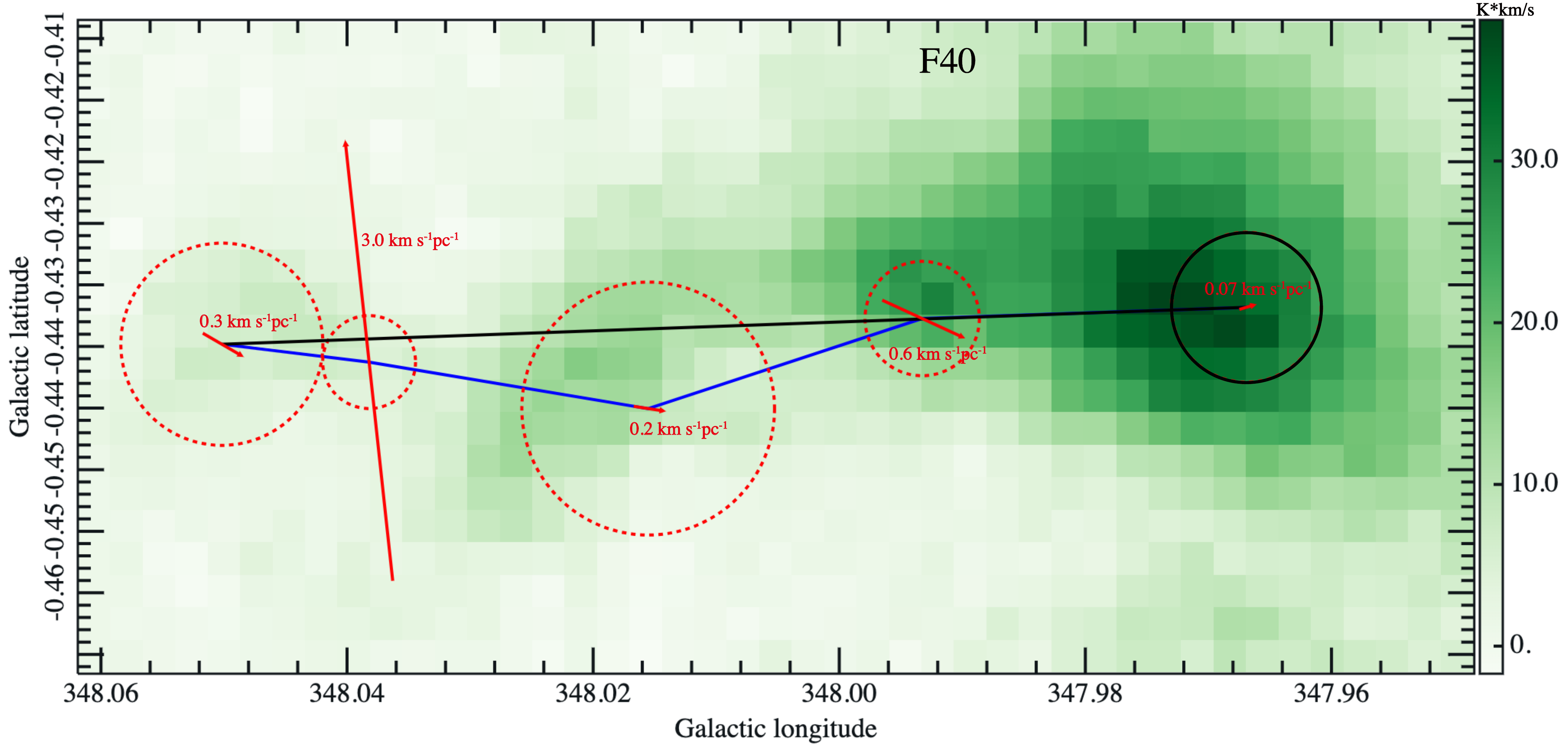}
		\includegraphics[width=0.45\textwidth]{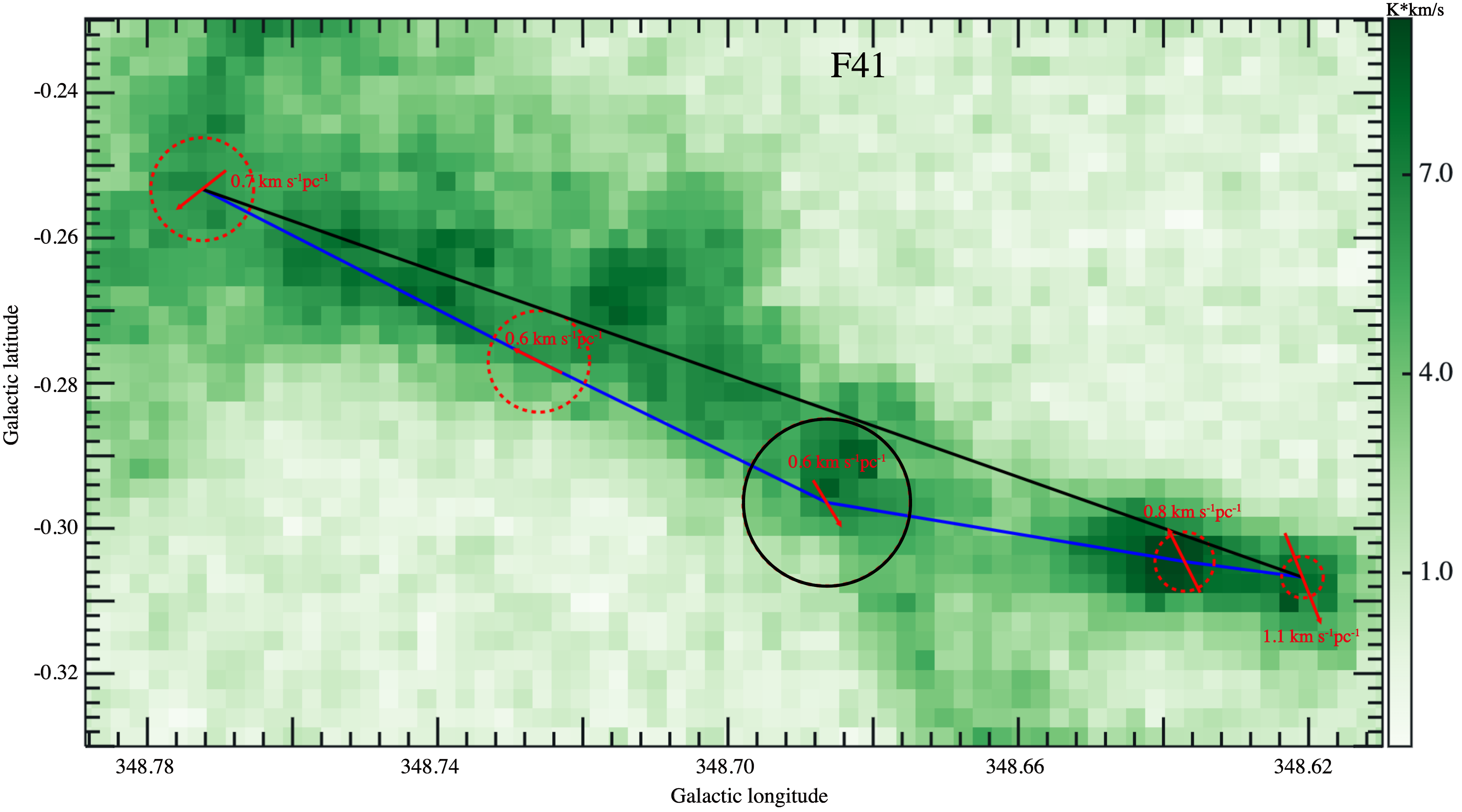}
	}%
	\caption{Clumps plotted on the integrated intensity map of $^{13}$CO (2-1). The circles show the size and location of the clumps. The red arrows with varying length represent the velocity gradients. The lengths of red arrows reflect the values of the measured velocity gradients. The blue solid lines are the the fitted filaments, whose orientations are represent by the black solid lines (see Section~\ref{sec:fila}). The black solid circles are the most massive clumps in each target filament. \label{fig:direction}}         
\end{figure*}

The filaments were identified from Hi-GAL clumps~\citep{Elia:21} using the Minimum Spanning Tree (MST) algorithm developed by~\citet{Wang:16}. The specialized MST algorithm is applied to identify filaments in Position-Position-Velocity space, implying that clumps are only clustered as a filament when they are in close proximity with comparable velocities. In Figure~\ref{fig:direction}, the clumps are marked with circles and linked by minimizing the total length of connecting edges. The edges represented by blue solid lines are straight lines joining the clumps, and the lengths of the edges indicate the distance between each pair of clumps. The fitted filaments are marked by end-to-end lines connecting two clumps at the filament tips, shown as the black solid lines in Figure~\ref{fig:direction}. Details of the identified linear filaments can be found in~\citet{Wang:24}. Orientations ($\theta_{f}$) of the target filaments were determined by using the positions of the clumps at the two tips, with measurements taken clockwise from the East. 

\subsection{Velocity Gradient Fitting} \label{sec:gradient}
For most filaments, the distances between clumps are consistent with the findings reported by~\citet{Elia:21}. 
According to~\citet{Wang:24}, there is no evident selection bias in the distance distributions of the target filaments, supporting the assumption that all clumps within a target filament are equidistant. Consequently the related parameters of the clumps reported in~\citet{Elia:21} 
are scaled to derive the velocity gradient, 
the specific angular momentum ($J/M$), and the ratio ($\beta$) between 
the rotational energy and gravitational energy. 
The scaled properties of clumps in target filaments 
are listed in Table~\ref{table:infclump}.  
Clumps, which appear as localized increases in column density within clouds, are generally assumed to be spherical in shape with rigid-body rotation, producing a linear gradient, $\mathbf\nabla{}v_\mathrm{LSR}$. 
In the local standard of rest (LSR) velocity field, 
the linear gradient aligns perpendicular to the rotation axis. 

\begin{table*}
	\caption{The information of clumps in filaments.}\label{table:infclump}
	\centering
	\footnotesize
	\begin{tabular}{cccccccc}    
		\hline
		\hline
		Filaments & $\ell$ & b & Distance & FWHM$_{250}$ & Diameter & Mass & $v_{0}$ \\  
		{} & (degree) & (degree) & (kpc) & $''$ & (pc) & (M$_{\odot}$) & (km~s$^{-1}$) \\
		(1) & (2) & (3) & (4) & (5) & (6) & (7) & (8) \\
		\hline
		{} & 11.89 & -0.40 & 12.0  & 15.8 & 0.9 & 1.9E+3(4.8) & 47.2 \\        
		{} & 11.91 & -0.40 & 12.0  & 34.5 & 2.0 & 4.7E+2(0.1) & 46.4 \\        
		F2 & 11.95 & -0.40 & 12.0  & 29.7 & 1.7 & 1.1E+3(2.0) & 45.9 \\        
		{} & 12.0  & -0.42 & 12.0  & 20.9 & 1.2 & 3.0E+2(0.6) & 44.2 \\        
		{} & 12.05 & -0.42 & 12.0  & 41.3 & 2.4 & 2.3E+3(3.9E+1) & 42.9 \\ \hline            		              
		{} & 14.65 & 0.26  & 2.7   & 17.4 & 0.2 & 1.7E+1(1.7E+1) & 26.9 \\        
		{} & 14.64 & 0.30 & 2.7    & 15.9 & 0.2 & 3.8E+1(1.3) & 26.4 \\        
		F3 & 14.63 & 0.31 & 2.7    & 16.1 & 0.2 &4.0E+2(1.5E+2) & 25.6 \\        
		{} & 14.62 & 0.33 & 2.7    & 20.4 & 0.3 & 2.5E+2(6.7) & 27.0 \\        
		{} & 14.61 & 0.38 & 2.7    & 33.7 & 0.4 & 5.2E+1(1.5) & 27.0 \\ \hline 		
		{} & 328.84 & 0.04 & 12.1  & 28.5 & 1.7 & 5.5E+2(2.7) & -33.8 \\       
		{} & 328.83 & 0.05 & 12.1  & 12.5 & 0.7 & 5.7E+2(5.0E+2) & -33.8 \\       
		F34 & 328.82 & 0.06 & 12.1 & 12.7 & 0.7 & 4.8E+2(4.1E+2) & -33.0 \\       
		{} & 328.81 & 0.07 & 12.1  & 14.2 & 0.8 & 1.8E+3(1.9E+2) & -32.5 \\       
		{} & 328.81 & 0.07 & 12.1  & 13.3 & 0.8 & 2.7E+3(5.8E+2) & -31.2 \\ \hline
		{} & 338.47 & 0.21 & 2.8   & 12.8 & 0.2 & 9.0E+1(2.7E+1) & -35.6 \\       
		{} & 338.50 & 0.21 & 2.8   & 10.8 & 0.1 & 2.3E+2(2.8) & -36.6 \\       
		F35 & 338.53 & 0.21 & 2.8  & 28.4 & 0.4 & 2.8E+2(8.4E+2) & -38.2 \\      
		{} & 338.56 & 0.22 & 2.8   & 8.6  & 0.1 & 1.9E+2(7.9) & -35.5 \\       
		{} & 338.58 & 0.21& 2.8    & 29.8 & 0.4 & 1.3E+1(1.1E+1) & -36.8 \\ \hline		
		{} & 338.52 & -0.43 & 2.9  & 30.2 & 0.4 & 1.8E+2(8.3E+2) & -38.8 \\       
		{} & 338.55 & -0.42 & 2.9  & 22.7 & 0.3 & 3.2E+2(1.1E+2) & -38.3 \\       
		{} & 338.60 & -0.44 & 2.9  & 10.9 & 0.2 & 7.6E+1(1.8E+1) & -38.4 \\       
		{} & 338.65 & -0.45 & 2.9  & 15.3 & 0.2 & 3.1E+2(8.4E+1) & -39.3 \\       
		{} & 338.66 & -0.45 & 2.9  & 14.8 & 0.2 & 5.0E+1(7.3) & -38.6 \\       
		F36 & 338.69 & -0.48 & 2.9 & 28.0 & 0.4 & 1.9E+2(4.0E+1) & -39.1 \\       
		{} & 338.72 & -0.46 & 2.9  & 12.5 & 0.2 & 5.8E+1(1.5E+1) & -39.5 \\       
		{} & 338.77 & -0.46 & 2.9  & 27.5 & 0.4 & 1.6E+3(3.7E+3) & -38.8 \\       
		{} & 338.82 & -0.45 & 2.9  & 22.0 & 0.3 & 2.0E+2(7.9E+2) & -40.7 \\       
		{} & 338.82 & -0.47 & 2.9  & 22.2 & 0.3 & 1.9E+2(1.9E+3) & -37.9 \\       
		{} & 338.87 & -0.48 & 2.9  & 14.1 & 0.2 & 3.7E+2(1.0E+1) & -37.2 \\ \hline 		
		{} & 346.02 & -0.02 & 10.8 & 14.2 & 0.7 & 3.2E+2(5.0E+2) & -80.9 \\       
		{} & 345.88 & -0.01& 10.8  & 19.9 & 1.0 & 1.1E+3(2.4E+3) & -78.8 \\       
		{} & 345.92 &  0. & 10.8   & 15.2 & 0.8 & 1.3E+3(3.1E+2) & -79.6 \\       
		F39 & 345.93 &  0.& 10.8   & 27.2 & 1.4 & 2.4E+3(1.4E+2) & -79.8 \\       
		{} & 345.98 & -0.02 & 10.8 & 10.2 & 0.5 & 1.6E+3(1.6E+2) & -80.9 \\       
		{} & 346.06 & -0.03 & 10.8 & 25.0 & 1.3 & 4.9E+3(2.6E+3) & -81.6 \\ \hline		
		{} & 347.97 & -0.43 & 6.0  & 21.9 & 0.6 & 4.7E+3(8.6E+2) & -94.9 \\       
		{} & 347.99 & -0.43 & 6.0  & 16.7 & 0.5 & 2.9E+2(2.3E+2) & -94.2 \\       
		F40 & 348.02 & -0.44 & 6.0 & 37.0 & 1.1 & 5.0E+2(1.5E+3) & -93.5 \\       
		{} & 348.04 & -0.44 & 6.0  & 13.6 & 0.4 & 2.7E+1(2.7E+1) & -95.6 \\       
		{} & 348.05 & -0.43 & 6.0  & 29.6 & 0.9 & 8.0E+2(3.9E+2) & -95.2 \\ \hline		
		{} & 348.62 & -0.32 & 2.2  & 10.4 & 0.1 & 2.6E+1(3.0) & -19.1 \\
		{} & 348.64 & -0.31 & 2.2  & 14.7 & 0.2 & 5.4E+1(2.3E+2) & -19.1 \\
		F41 & 348.69 & -0.31 & 2.2 & 41.5 & 0.4 & 1.0E+2(3.1E+3) & -19.2 \\
		{} & 348.73 & -0.29 & 2.2  & 25.1 & 0.3 & 7.0E+1(2.1E+3) & -19.4 \\
		{} & 348.77 & -0.26 & 2.2  & 25.5 & 0.3 & 4.5E+1(9.1E+1) & -19.8 \\ 
		\hline
		\multicolumn{8}{l}{Columns are (1) filament name, (2) galactic longitude, (3) galactic latitude, 
			(4) distance of the clump, adopted from~\citet{Wang:24},} \\
			\multicolumn{8}{l}{(5) the scaled size of the clump as estimated by `CuTEx'~\citep{Elia:17} in the 250 $\mu$m band, }\\
			\multicolumn{8}{l}{(6) the scaled linear diameter of the clump, obtained combining FWHM$_{250}$ and distance,} \\
			\multicolumn{8}{l}{(7) the scaled mass and its uncertainty of the clump, (8) velocity of the clump, assigned by~\citet{Mege:21}.} \\
	\end{tabular}
\end{table*} 

The velocity gradients were measured following the method 
described in~\citet{Goodman:93}, which fits the function
\begin{equation}
	v_{LSR} = v_0 + c_{1}{}\Delta{}l + c_{2}{}\Delta{}b 
	\label{equ1}.
\end{equation}
Here, $v_{LSR}$ represents an intensity weighted average velocity 
along the line of sight and $v_{0}$ is velocities of clumps (see column (8) in Table~\ref{table:infclump}).
$\Delta{}l$ and $\Delta{}b$ are the offsets from the center positions 
(listed in column (2) and (3) of Table~\ref{table:infclump}) of clumps in the Galactic longitude and Galactic latitude in radians, respectively, 
while $c_{1}$ and $c_{2}$ are the projections of the gradient per radian onto the $l$ and $b$ axes, respectively. 
Finally, the magnitude of the velocity gradient is defined by
\begin{equation}
	{\cal{}G}= |\mathbf\nabla
	v_\mathrm{LSR}|=\frac{(c_{1}^{2}+c_{2}^{2})^{1/2}}{D}
	\label{equ2},
\end{equation}
where $D$ is the distance (column (4) in Table~\ref{table:infclump}) of clumps.
Its orientation (i.e., the orientation of the increasing velocity) measured
east of north is given by
\begin{equation}
	\theta_{\cal{}G}= \tan{}\frac{c_{1}}{c_{2}}
	\label{equ3}.
\end{equation}

\begin{figure*}
	\subfigure[]{
		\centering
		\includegraphics[width=0.48\textwidth]{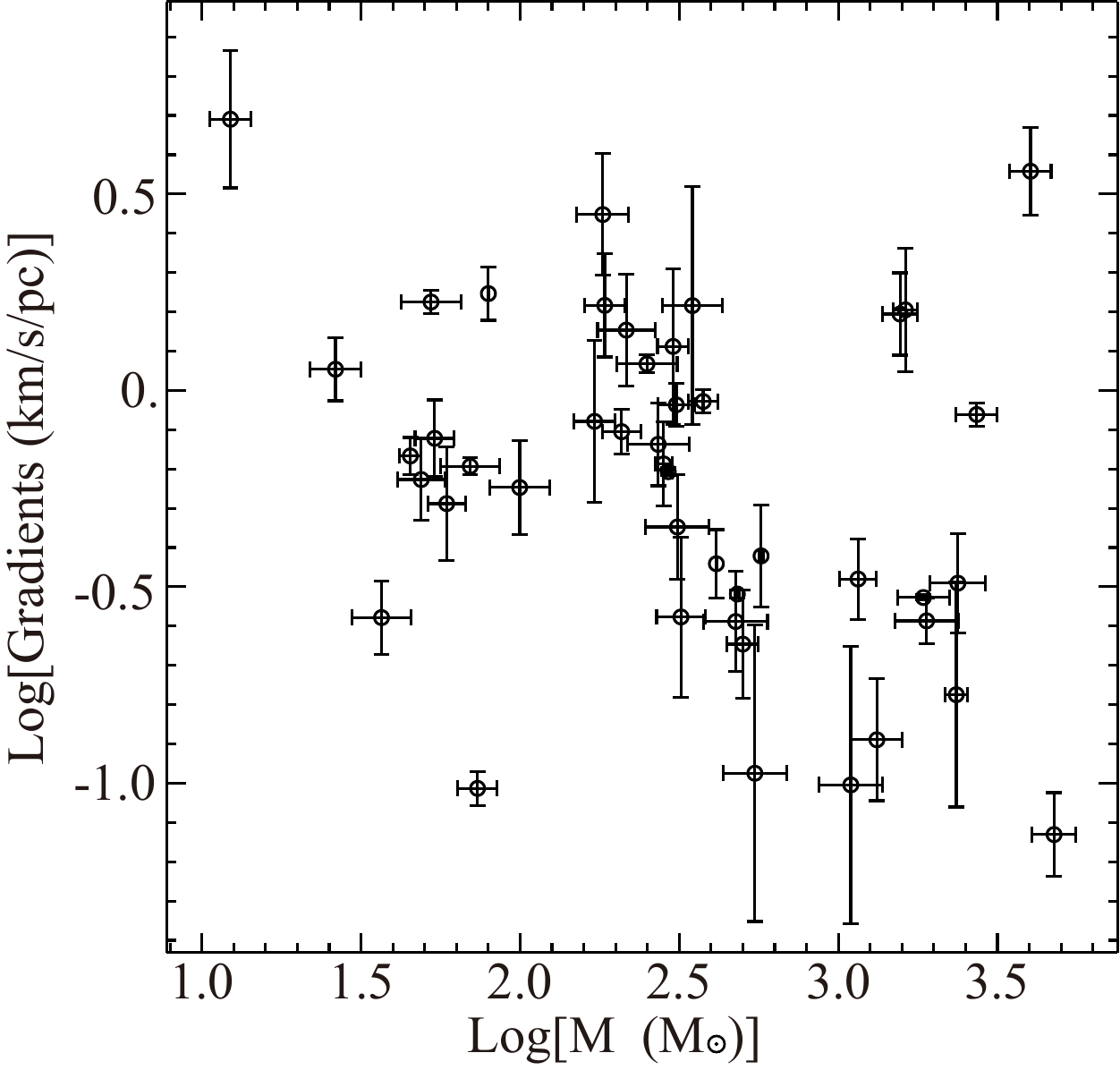}
	}%
	\subfigure[]{
		\centering
		\includegraphics[width=0.48\textwidth]{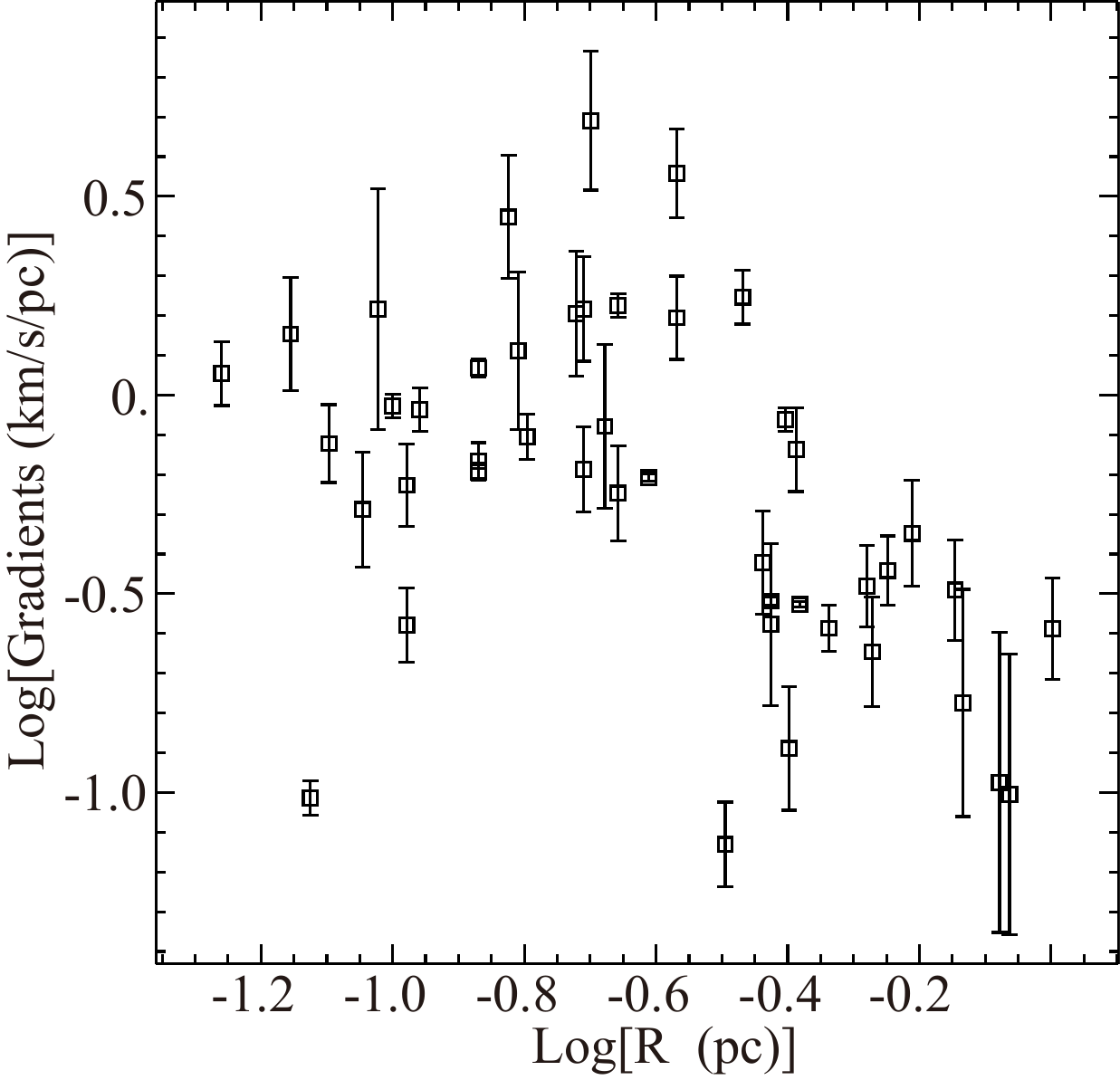}
	}%
	\caption{Velocity gradients plotted with (a) clump mass and (b) clump radius. 
		\label{fig:gradient}}           
\end{figure*}

Based on the spherical shape of clumps, the velocity field of the HiGAL clumps 
within the target linear filaments was fitted using a least-squares method 
as outlined in Equation (\ref{equ1}) from the $^{13}$CO (2-1) images. 
The magnitude of the gradient ($\cal{}G$), 
its direction ($\theta_{\cal{}G}$), 
and their errors were then calculated based on the resulted 
$c_{1}$ and $c_{2}$ of the least-squares fit. 
According to the confidence-level simulation results in~\citet{Goodman:93}, 
the velocity gradient of the target clumps can be reliably fitted 
when they contain at least nine spatial pixels in the $^{13}$CO (2-1) images. 
One clump in each of the two filaments 
(F35 and F39) was not well fitted. 
The derived $\cal{}G$ values distribute between 7.0~$\times$~10$^{-2}$\,km~s$^{-1}$pc$^{-1}$ and 4.8\,km~s$^{-1}$pc$^{-1}$, 
listed in column (4) of Table~\ref{table:dypa}. 
Its direction, $\theta_{\cal{}G}$, are presented in column (5) of Table~\ref{table:dypa}. 
Figure~\ref{fig:gradient} shows the distribution of the obtained $\cal{}G$ values 
compared with the clump radius ($R$) and mass ($M$). 
One could note that $\cal{}G$ tends to decrease as $R$ and $M$ increase.

\section{Results} \label{sec:res}

\begin{table*}
	\caption{Derived clump properties}\label{table:dypa}
	\centering
	\footnotesize
	\begin{tabular}{cccccccc}     
		\hline
		\hline
		Filaments & $\ell$ & b & $\cal G$ & $\theta_{\cal G}$ & J/M & $\beta$ & $\vert{}\theta_{f} - \theta_{\cal G}{}\vert$\\  
		{} & (degree) & (degree) & (km\,s$^{-1}$\,pc$^{-1}$) & (degree E of N) & (pc\,km\,s$^{-1}$) & {} & (degree) \\
		(1) & (2) & (3) & (4) & (5) & (6) & (7) & (8) \\
		\hline
		{} & 11.89 & -0.40& 0.3 $\pm$ 0.1 &   56.5 $\pm$ 25.8  & 1.5~$\times$~10$^{-2}$ $\pm$ 8.9~$\times$~10$^{-3}$ & 1.1~$\times$~10$^{-3}$ $\pm$ 3.9~$\times$~10$^{-4}$ & 43.5 $\pm$ 25.8 \\		
		{} & 11.91 & -0.40 & 0.3 $\pm$ 0.03 &   90.7 $\pm$  6.1  & 7.2~$\times$~10$^{-2}$ $\pm$ 9.1~$\times$~10$^{-3}$ & 4.5~$\times$~10$^{-2}$ $\pm$ 7.4~$\times$~10$^{-4}$ & 10.3 $\pm$  6.1 \\		
		F2 & 11.95 & -0.40 & 0.1 $\pm$ 0.04 &  155.6 $\pm$ 12.9  & 2.0~$\times$~10$^{-2}$ $\pm$ 7.2~$\times$~10$^{-3}$ & 1.8~$\times$~10$^{-3}$ $\pm$ 2.3~$\times$~10$^{-4}$ & 55.7 $\pm$ 12.9 \\		
		{} & 12.0 & -0.42 & 0.5 $\pm$ 0.06 & -100.0 $\pm$  5.5  & 4.7~$\times$~10$^{-2}$ $\pm$ 6.2~$\times$~10$^{-3}$ & 4.8~$\times$~10$^{-2}$ $\pm$ 8.5~$\times$~10$^{-4}$ & 20.9 $\pm$  5.5 \\		
		{} & 12.05 & -0.42 & 0.2 $\pm$ 0.03 &  -15.2 $\pm$  8.1  & 3.4~$\times$~10$^{-2}$ $\pm$ 3.6~$\times$~10$^{-3}$ & 4.3~$\times$~10$^{-2}$ $\pm$ 4.8~$\times$~10$^{-4}$ & 64.8 $\pm$  8.1  \\  \hline
		{} & 14.65 & 0.26 & 1.8 $\pm$ 0.6 &  138.3 $\pm$ 4.7 & 3.2~$\times$~10$^{-2}$ $\pm$ 7.5~$\times$~10$^{-3}$   & 1.8~$\times$~10$^{-2}$   $\pm$ 1.4~$\times$~10$^{-3}$ & 13.6  $\pm$ 4.7 \\       
		{} & 14.63 & 0.30 & 0.3 $\pm$ 0.02 &  -16.2 $\pm$ 6.0 & 7.8~$\times$~10$^{-4}$ $\pm$ 4.2~$\times$~10$^{-5}$ & 7.0~$\times$~10$^{-4}$ $\pm$ 9.9~$\times$~10$^{-5}$  & 11.9 $\pm$ 6.0 \\          
		F3 & 14.63  & 0.31 & 0.9 $\pm$ 0.3  &  23.1 $\pm$ 30.4 & 2.6~$\times$~10$^{-2}$ $\pm$ 7.7~$\times$~10$^{-4}$ & 7.4~$\times$~10$^{-4}$ $\pm$ 6.6~$\times$~10$^{-5}$  & 47.3 $\pm$ 30.4 \\       
		{} & 14.62  & 0.33 & 1.2 $\pm$ 0.3  & -56.7 $\pm$ 11.0 & 5.8~$\times$~10$^{-3}$ $\pm$ 3.1~$\times$~10$^{-4}$ & 4.2~$\times$~10$^{-3}$ $\pm$ 2.1~$\times$~10$^{-4}$  & 30.6 $\pm$ 11.0 \\        
		{} & 14.61  & 0.38 & 1.7 $\pm$ 0.5  & 123.0 $\pm$ 11.3 & 2.2~$\times$~10$^{-3}$ $\pm$ 6.4~$\times$~10$^{-4}$ & 1.3~$\times$~10$^{-1}$ $\pm$ 1.5~$\times$~10$^{-2}$  & 30.9 $\pm$ 11.3 \\ \hline
		{} & 328.84 & 0.04 & 0.1 $\pm$ 0.04 & 147.3 $\pm$ 16.1 & 2.0~$\times$~10$^{-3}$ $\pm$ 7.7~$\times$~10$^{-4}$ & 3.8~$\times$~10$^{-3}$ $\pm$ 5.4~$\times$~10$^{-4}$  & 4.5 $\pm$ 16.1 \\         
		{} & 328.83 & 0.05 & 0.4 $\pm$ 0.1  &  28.3 $\pm$ 14.1 & 1.4~$\times$~10$^{-3}$ $\pm$ 5.4~$\times$~10$^{-4}$ & 3.9~$\times$~10$^{-3}$ $\pm$ 6.0~$\times$~10$^{-4}$  &  65.5 $\pm$ 14.1  \\     
		F34 & 328.82 & 0.06 & 0.3 $\pm$ 0.2  & 162.3 $\pm$ 2.9  & 1.2~$\times$~10$^{-3}$ $\pm$ 7.8~$\times$~10$^{-4}$ & 3.2~$\times$~10$^{-3}$ $\pm$ 1.4~$\times$~10$^{-4}$  & 21.5 $\pm$ 2.9 \\        
		{} & 328.81 & 0.07 & 0.3 $\pm$ 0.2  & 126.4 $\pm$ 30.4 & 1.4~$\times$~10$^{-3}$ $\pm$ 1.1~$\times$~10$^{-3}$ & 1.1~$\times$~10$^{-3}$ $\pm$ 6.2~$\times$~10$^{-4}$  & 16.3 $\pm$ 30.4 \\        
		{} & 328.81 & 0.07 & 0.9 $\pm$ 0.04 &  47.5 $\pm$ 7.6  & 3.7~$\times$~10$^{-3}$ $\pm$ 1.4~$\times$~10$^{-3}$ & 5.4~$\times$~10$^{-3}$ $\pm$ 7.5~$\times$~10$^{-4}$  & 79.8 $\pm$ 7.6 \\ \hline  				
		{}  & 338.47 & 0.21 & $-$ & $-$ & $-$ & $-$ & $-$ \\                                                                   
		{}  & 338.50 & 0.21  & 4.8 $\pm$ 0.7  & -5.0 $\pm$ 23.2   & 9.7~$\times$~10$^{-4}$ $\pm$ 6.5~$\times$~10$^{-4}$ & 5.7~$\times$~10$^{-2}$ $\pm$ 1.9~$\times$~10$^{-3}$  & 79.3 $\pm$ 23.2 \\        
	F35  & 338.53 & 0.21  & 3.7 $\pm$ 0.4  & -72.4 $\pm$ 14.6  & 4.9~$\times$~10$^{-3}$ $\pm$ 5.4~$\times$~10$^{-4}$ & 6.6~$\times$~10$^{-3}$ $\pm$ 8.2~$\times$~10$^{-5}$  & 30.9 $\pm$ 14.6 \\       
	{}  & 338.56 & 0.22  & 3.0 $\pm$ 0.2  & -153.6 $\pm$ 43.5 & 6.8~$\times$~10$^{-3}$ $\pm$ 7.3~$\times$~10$^{-4}$ & 3.5~$\times$~10$^{-3}$ $\pm$ 4.1~$\times$~10$^{-4}$  & 75.3 $\pm$ 43.5 \\        
	{}  & 338.58 & 0.21  & 1.4 $\pm$ 0.2  &  72.8 $\pm$  6.2  & 2.5~$\times$~10$^{-3}$ $\pm$ 4.0~$\times$~10$^{-4}$ & 3.1~$\times$~10$^{-3}$ $\pm$ 6.3~$\times$~10$^{-5}$  & 6.0 $\pm$ 6.2 \\ \hline   
	{}  & 338.52 & -0.43 & 0.8 $\pm$ 0.2  & 6.9 $\pm$ 8.4     & 1.0~$\times$~10$^{-2}$ $\pm$ 2.1~$\times$~10$^{-3}$ & 1.2~$\times$~10$^{-2}$ $\pm$ 5.1~$\times$~10$^{-4}$  & 89.3 $\pm$ 8.4  \\        
	{}  & 338.55 & -0.42 & 1.3 $\pm$ 0.3  &   83.4 $\pm$  7.9 & 8.5~$\times$~10$^{-3}$ $\pm$ 1.7~$\times$~10$^{-3}$ & 6.5~$\times$~10$^{-3}$ $\pm$ 2.6~$\times$~10$^{-4}$  & 14.2 $\pm$ 7.9 \\         
	{}  & 338.60 & -0.44 & 0.1 $\pm$ 0.04 & 46.9 $\pm$ 18.5   & 1.5~$\times$~10$^{-3}$ $\pm$ 3.7~$\times$~10$^{-4}$ & 1.7~$\times$~10$^{-4}$ $\pm$ 1.1~$\times$~10$^{-6}$  & 50.7 $\pm$ 18.5  \\       
	{}  & 338.65 & -0.45 & 0.9 $\pm$ 0.5  & 101.5 $\pm$ 2.8   & 3.1~$\times$~10$^{-3}$ $\pm$ 1.7~$\times$~10$^{-4}$ & 1.1~$\times$~10$^{-3}$ $\pm$ 3.4~$\times$~10$^{-4}$  & 3.9 $\pm$ 2.8 \\          
	{}  & 338.66 & -0.45 & 0.6 $\pm$ 0.06 &  135.4 $\pm$ 42.2 & 1.8~$\times$~10$^{-3}$ $\pm$ 1.9~$\times$~10$^{-4}$ & 2.6~$\times$~10$^{-3}$ $\pm$ 2.9~$\times$~10$^{-4}$  & 41.9 $\pm$ 42.2  \\       
	F36  & 338.69 & -0.48 & 1.6 $\pm$ 0.2  &  -77.9 $\pm$  5.2 & 1.7~$\times$~10$^{-2}$ $\pm$ 2.3~$\times$~10$^{-3}$ & 3.4~$\times$~10$^{-2}$ $\pm$ 5.9~$\times$~10$^{-4}$  & 4.5 $\pm$ 5.2 \\         
	{}  & 338.72 & -0.46 & 0.5 $\pm$ 0.07 & -177.4 $\pm$ 30.0 & 1.1~$\times$~10$^{-3}$ $\pm$ 1.7~$\times$~10$^{-4}$ & 1.0~$\times$~10$^{-3}$ $\pm$ 2.2~$\times$~10$^{-4}$  & 80.0 $\pm$ 30.0 \\        
	{}  & 338.77 & -0.46 & 1.6 $\pm$ 0.3  &  162.8 $\pm$  6.4 & 1.6~$\times$~10$^{-2}$ $\pm$ 2.5~$\times$~10$^{-3}$ & 3.4~$\times$~10$^{-3}$ $\pm$ 8.6~$\times$~10$^{-5}$  & 65.3 $\pm$  6.4 \\        
	{}  & 338.82 & -0.45 & 0.8 $\pm$ 0.4  &   61.1 $\pm$ 23.4 & 5.5~$\times$~10$^{-3}$ $\pm$ 3.1~$\times$~10$^{-4}$ & 3.8~$\times$~10$^{-3}$ $\pm$ 1.2~$\times$~10$^{-4}$  & 40.4 $\pm$ 23.4  \\       
	{}  & 338.82 & -0.47 & 2.7 $\pm$ 0.4  & -168.4 $\pm$  6.2 & 1.7~$\times$~10$^{-2}$ $\pm$ 2.7~$\times$~10$^{-3}$ & 4.6~$\times$~10$^{-2}$ $\pm$ 1.1~$\times$~10$^{-3}$  & 86.0 $\pm$  6.2 \\        
	{}  & 338.87 & -0.48 & 1.6 $\pm$ 0.5  &  -53.1 $\pm$ 12.1 & 4.1~$\times$~10$^{-3}$ $\pm$ 1.2~$\times$~10$^{-4}$ & 2.1~$\times$~10$^{-3}$ $\pm$ 1.9~$\times$~10$^{-4}$  & 31.4 $\pm$ 12.1 \\ \hline
	{}  & 345.88 & -0.01 & 0.3 $\pm$ 0.1   &  115.8 $\pm$ 17.9 & 1.0~$\times$~10$^{-2}$ $\pm$ 2.1~$\times$~10$^{-3}$ & 3.7~$\times$~10$^{-3}$ $\pm$ 1.5~$\times$~10$^{-4}$ & 20.8 $\pm$ 17.9 \\    
	{}  & 345.92 &  0. & 0.1 $\pm$ 0.02    & -114.8 $\pm$  6.2 & 2.5~$\times$~10$^{-2}$ $\pm$ 1.0~$\times$~10$^{-3}$ & 4.3~$\times$~10$^{-3}$ $\pm$ 7.5~$\times$~10$^{-4}$ & 31.9 $\pm$  6.2 \\    
	F39  & 345.93 &  0. & 0.3 $\pm$ 0.04    & -24.4 $\pm$  16.6 & 5.7~$\times$~10$^{-3}$ $\pm$ 8.8~$\times$~10$^{-4}$ & 2.6~$\times$~10$^{-4}$ $\pm$ 6.1~$\times$~10$^{-5}$ & 61.6 $\pm$  16.6 \\  
	{}  & 345.98 & -0.02 & 1.6 $\pm$ 0.2   &    2.2 $\pm$  5.4 & 4.5~$\times$~10$^{-2}$ $\pm$ 6.3~$\times$~10$^{-3}$ & 5.1~$\times$~10$^{-3}$ $\pm$ 9.9~$\times$~10$^{-4}$ & 85.2 $\pm$  5.4 \\    
	{}  & 346.02 & -0.012 & 0.3 $\pm$ 0.06 &  1.3 $\pm$  7.5   & 3.1~$\times$~10$^{-2}$ $\pm$ 3.3~$\times$~10$^{-3}$ & 9.8~$\times$~10$^{-3}$ $\pm$ 1.1~$\times$~10$^{-4}$ & 79.2 $\pm$  7.5 \\    
	{}  & 346.06 & -0.03 & $-$ & $-$ & $-$ & $-$ & $-$ \\ \hline 
	{}  & 347.97 & -0.43 & 0.07$\pm$ 0.008 &  -73.5 $\pm$ 12.8 & 2.1~$\times$~10$^{-3}$ $\pm$ 2.2~$\times$~10$^{-4}$ & 1.1~$\times$~10$^{-5}$ $\pm$ 1.4~$\times$~10$^{-6}$& 14.4 $\pm$ 12.8 \\ 
	{}  & 347.99 & -0.43 & 0.6 $\pm$ 0.1   & -114.7 $\pm$  6.8 & 1.0~$\times$~10$^{-2}$ $\pm$ 1.7~$\times$~10$^{-3}$ & 6.1~$\times$~10$^{-3}$ $\pm$ 1.8~$\times$~10$^{-4}$& 26.8  $\pm$  6.8 \\ 
	F40 & 348.02 & -0.44 & 0.2 $\pm$ 0.03  &  -98.7 $\pm$  5.8 & 1.8~$\times$~10$^{-2}$ $\pm$ 2.4~$\times$~10$^{-3}$ & 4.9~$\times$~10$^{-3}$ $\pm$ 9.3~$\times$~10$^{-5}$&  10.8 $\pm$  5.8  \\ 
	{}  & 348.04 & -0.44 & 3.0 $\pm$ 0.4   & 6.2 $\pm$ 13.4  &   5.6~$\times$~10$^{-2}$ $\pm$ 2.6~$\times$~10$^{-3}$ & 4.9~$\times$~10$^{-2}$ $\pm$ 2.2~$\times$~10$^{-4}$& 85.9 $\pm$ 13.4  \\ 
	{}  & 348.05 & -0.43 & 0.3 $\pm$ 0.01  & -120.1 $\pm$ 12.5 & 2.5~$\times$~10$^{-2}$ $\pm$ 7.1~$\times$~10$^{-3}$ & 1.5~$\times$~10$^{-3}$ $\pm$ 1.2~$\times$~10$^{-4}$& 32.2 $\pm$ 12.5 \\ \hline 
	{}  & 348.62 & -0.32 & 1.1 $\pm$ 0.3   & -158.36 $\pm$ 2.7 & 9.4~$\times$~10$^{-4}$ $\pm$ 2.2~$\times$~10$^{-5}$ & 2.5~$\times$~10$^{-3}$ $\pm$ 1.4~$\times$~10$^{-4}$& 52.9 $\pm$ 2.7 \\  
	{}  & 348.64 & -0.31 & 0.8 $\pm$ 0.07  &   26.7 $\pm$ 23.3 & 1.3~$\times$~10$^{-3}$ $\pm$ 8.4~$\times$~10$^{-4}$ & 1.7~$\times$~10$^{-3}$ $\pm$ 6.8~$\times$~10$^{-4}$& 43.9 $\pm$ 23.3 \\  
	F41 & 348.69 & -0.31 & 0.6 $\pm$ 0.07  & -148.8 $\pm$ 34.8 & 7.5~$\times$~10$^{-3}$ $\pm$ 9.0~$\times$~10$^{-4}$ & 1.1~$\times$~10$^{-2}$ $\pm$ 1.6~$\times$~10$^{-4}$& 43.4 $\pm$ 34.8 \\ 
	{}  & 348.73 & -0.29 & 0.6 $\pm$ 0.2   &   63.3 $\pm$  2.2 & 3.2~$\times$~10$^{-3}$ $\pm$ 8.3~$\times$~10$^{-4}$ & 4.6~$\times$~10$^{-3}$ $\pm$ 3.1~$\times$~10$^{-4}$& 7.3 $\pm$  2.2  \\ 
	{}  & 348.77 & -0.26 & 0.7 $\pm$ 0.2   &  128.3 $\pm$ 14.3 & 3.4~$\times$~10$^{-3}$ $\pm$ 1.2~$\times$~10$^{-3}$ & 8.0~$\times$~10$^{-3}$ $\pm$ 9.8~$\times$~10$^{-4}$& 60.7 $\pm$ 14.3  \\        
		\hline
		\multicolumn{8}{l}{Columns are (1) filament name, (2) galactic longitude, (3) galactic latitude, (4) fitted velocity gradient, (5) the direction of the fitted velocity gradient, } \\
	    \multicolumn{8}{l}{(6) the specific angular momentum,(7) the ratio between rotational energy and gravitational energy, (8) angles between filaments and velocity gradients. } \\
	    \multicolumn{8}{l}{$-$ represents no fitted. } 
	\end{tabular}
\end{table*} 

\subsection{Specific Angular Momentum and Energy} \label{sec:dynamic}
If molecular clouds are rotating, their current angular momentum could provide insights into their evolution processes. The specific angular momentum ($J/M$), which represents angular momentum per unit mass, is commonly used to compare angular momenta in different parts with comparable masses. For a spherical clump with a power-law density distribution, $\rho~\propto~r^{-\cal{}A}$, $J/M$ is computed as in~\citet{Xu:20b}:
\begin{equation}
	\frac{J}{M} = \frac{2(3-{\cal{}A})}{3(5-{\cal{}A})}{}{\cal{}G}{}R^{2}.
	\label{equ4}
\end{equation} 

\begin{figure*}
	\centering
	\subfigure[]{
		\centering
		\includegraphics[width=0.48\textwidth]{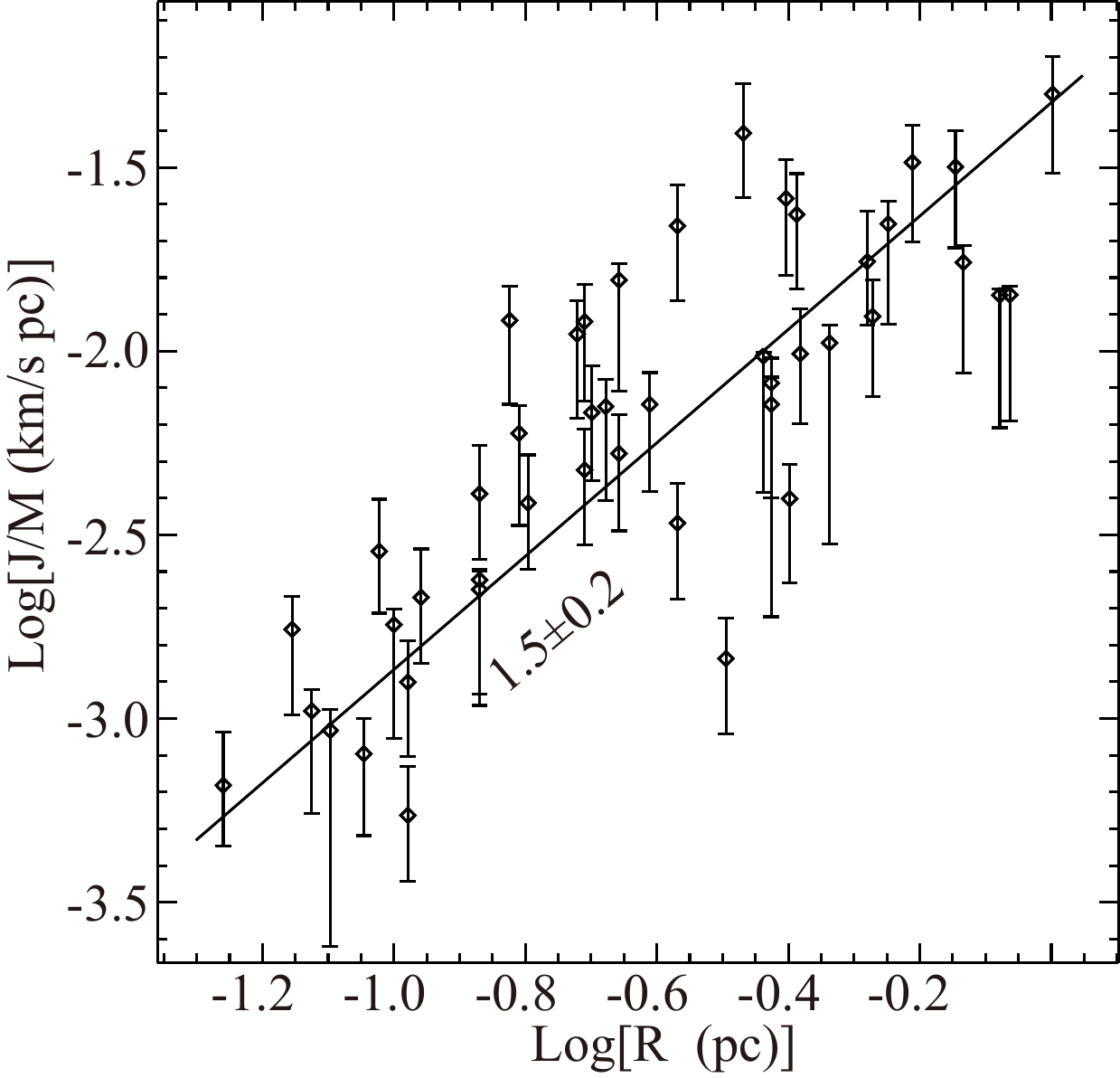}
	}%
	\subfigure[]{
		\centering
		\includegraphics[width=0.48\textwidth]{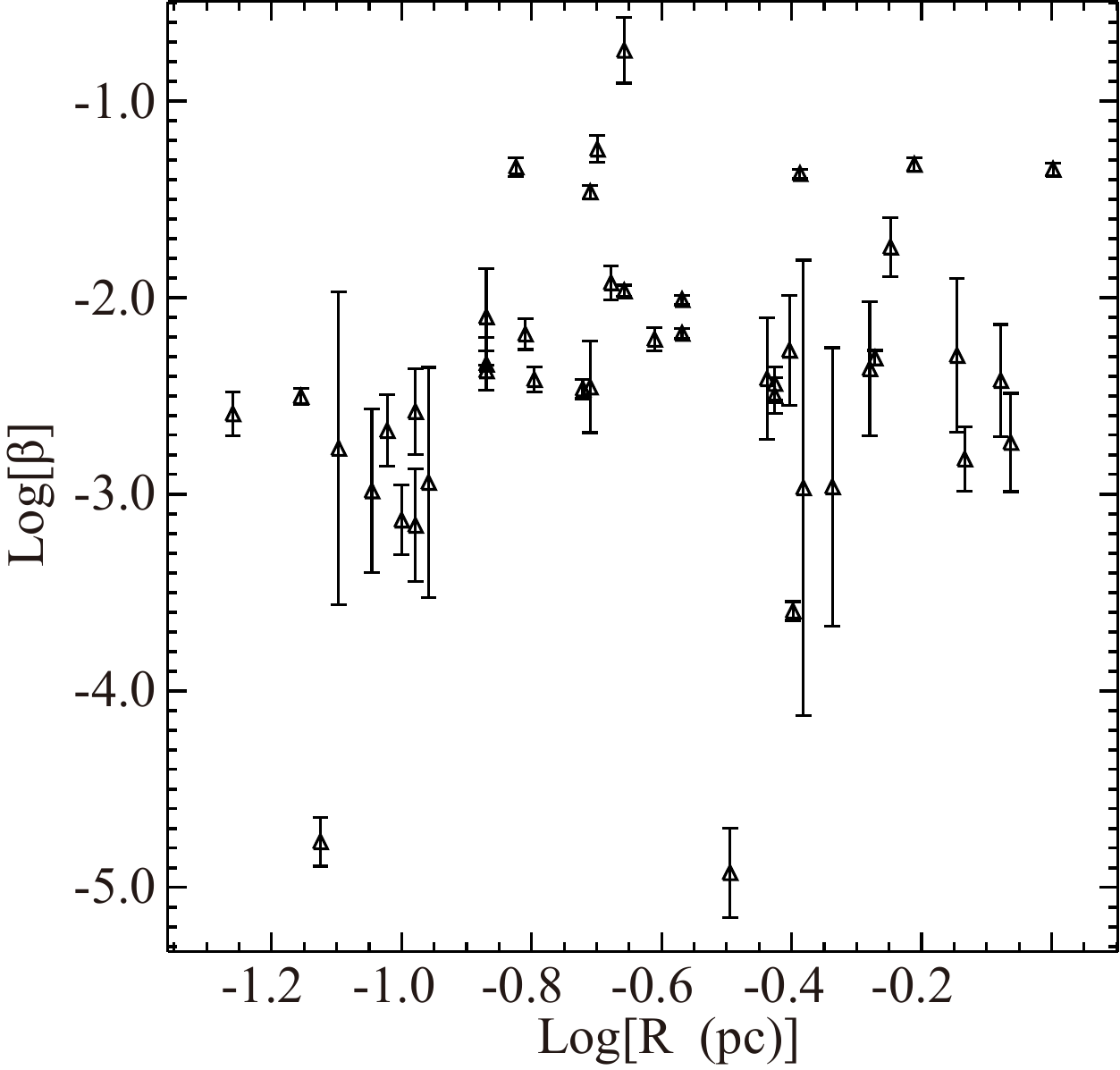}
	}%

	\subfigure[]{
	\centering
	\includegraphics[width=0.95\textwidth]{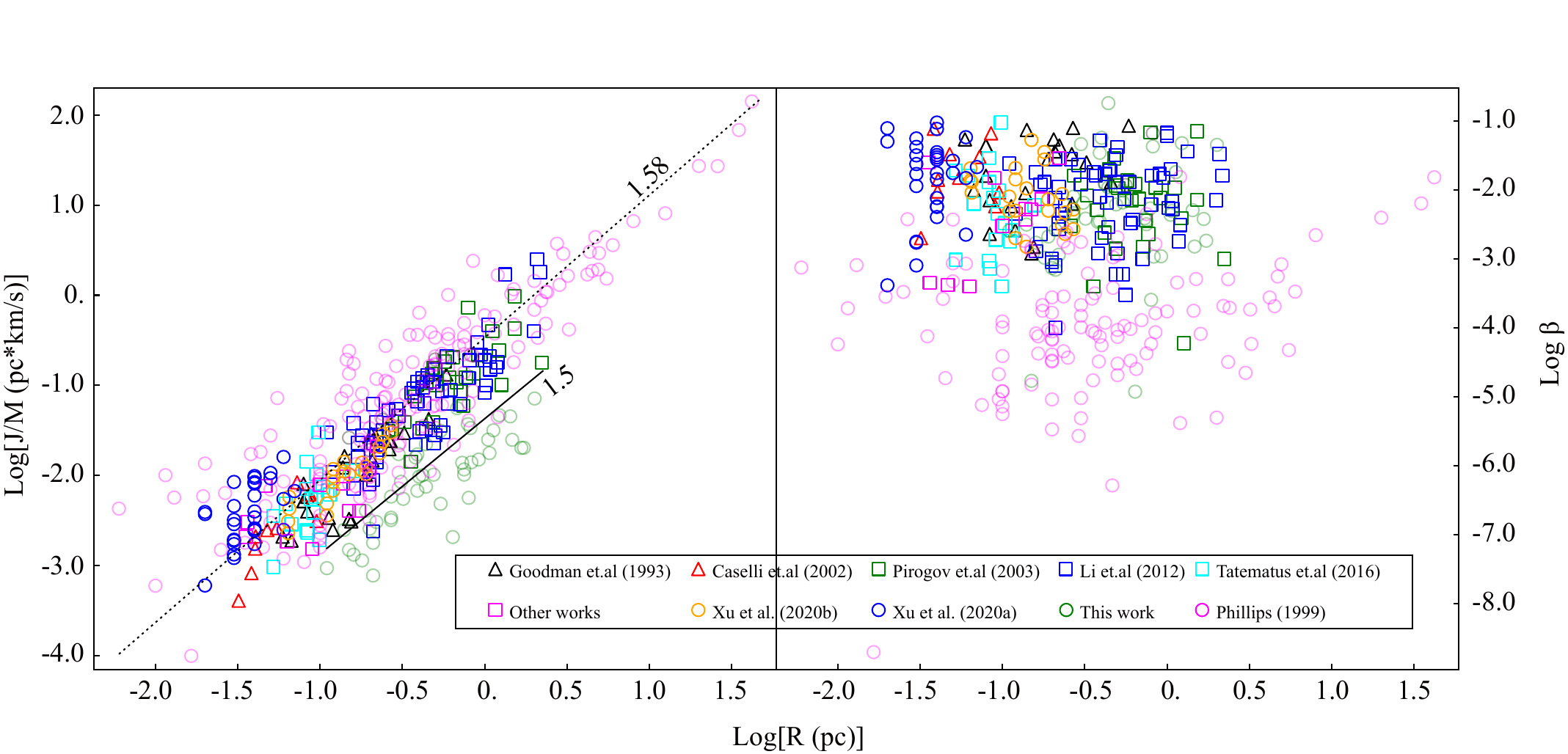}
}%

	\caption{(a) Specific angular momentum ($J/M$) and (b) the ratio ($\beta$) between rotational energy and gravitational energy plotted with clump radii ($R$). The best fitted power-law relation is obtained between $J/M$ and $R$. (c) The derived $J/M$ and $\beta$ in this work together with those previously observed. In the left panel, the dotted black line shows the best-fit slope, 1.58, for the whole clump sample. The solid black line exhibits the finding, 1.5, of this work.   
		\label{fig:moment}}           
\end{figure*}

When the density of a clump is uniform, $\cal{}A$ is set to 0. Here, $\cal{}A$ = 1.6~\citep{Bonnor:56} is adopted, resulting in $J/M$ being reduced by about 30\% compared to a uniform density. The assumption of spherical geometry introduces a systematic difference of 20\% in the results of specific angular momentum for elongated clumps when the rotation axis is considered parallel to the clump axis. The gradient direction, which in our assumption is a direction of clump rotation, is not always parallel to either of the clump axes. In fact, the angles appear to be random, suggesting that a spherical geometry is the most reasonable assumption for our analysis. The derived $J/M$ values are listed in column (6) of Table~\ref{table:dypa} and are plotted in Figure~\ref{fig:moment}(a). The derived $J/M$ values range from 8.0~$\times$~10$^{-4}$\,pc\,kms$^{-1}$ to 0.1\,pc\,kms$^{-1}$. We found a monotonic increase of $J/M$ as a function of clump size ($R$), following a power-law relation $J/M~\propto~R^{1.5\pm0.2}$, which is consistent with ~\citet{Goodman:93}. 

The ratio ($\beta$) between 
rotational energy (E$_{r}$) and gravitational energy (E$_{g}$) 
is used  to quantify the dynamical role of rotation, 
defined by~\citet{Xu:20b} as 
\begin{equation}
	\beta =  \frac{E_{r}}{E_{g}} = \frac{25{\cal{}G}^{2}R^{3}}{3(5-{\cal{}A}){}(5-2{\cal{}A})GM} 
	\label{equ5}
\end{equation}
In this equation, 
\begin{equation}	
	E_{r} = \frac{3-{\cal{}A}}{3(5-{\cal{}A})}MR^{2}{\cal{}G}^{2},
	E_{g} = -\frac{3-{\cal{}A}}{5-2{\cal{}A}}{}\frac{GM^{2}}{R}. \nonumber
\end{equation} 
The assumption of $\cal{}A$ = 1.6 \citep{Bonnor:56} leads to a 70\% reduction 
in E$_{r}$ and a 31\% increase in E$_{g}$, 
which cumulatively results in $\beta$ being two times lower than 
that of a uniform sphere.
The derived $\beta$ values range from 1.0~$\times$~10$^{-5}$ to 0.1, 
as given in column (7) of Table~\ref{table:dypa} 
and plotted in Figure~\ref{fig:moment}(b). 
The lack of variation in $\beta$ implies that it may remain 
constant regardless of $R$, a conclusion consistent with the 
results from~\citet{Goodman:93}. As explored in previous studies
~\citep[e.g.][]{Goodman:93,Curtis:11,Xu:20b}, 
the small values of $\beta$ signifies that rotation alone 
provides insignificant support against gravitational collapse. 
To facilitate clear comparison, the obtained values of $J/M$ and $\beta$ in this work were analyzed alongside those from previous woks, as illustrated in Figure~\ref{fig:moment}(c). This discussion can  be found in Section~\ref{sec:angular}.      

\subsection{Angles Between Filaments and Velocity Gradients} \label{sec:diffangle} 
To quantify the alignment between the velocity gradients of clumps 
and their natal filaments, the relative difference between the angles, $\vert{}\theta_{f} - \theta_{\cal{}G}{}\vert$, 
is derived through
\begin{equation}
	\vert{}\theta_{f} - \theta_{\cal{}G}{}\vert = MIN\{{}\vert{}\theta_{f} - \theta_{\cal{}G}{}\vert,{}\vert{}180-\vert{}\theta_{f} - \theta_{\cal{}G}{}\vert{}\vert{}\}\label{equ6},
\end{equation}
where `MIN' refers to the minimum angular difference 
between $\theta_{f}$ and $\theta_{\cal{}G}$. 
The derived $\vert{}\theta_{f} - \theta_{\cal{}G}{}\vert$ values range from 3.9 to 89.3, 
as presented in column (8) of Table~\ref{table:dypa}.

\begin{figure}
	\centering
	\includegraphics[width=0.48\textwidth]{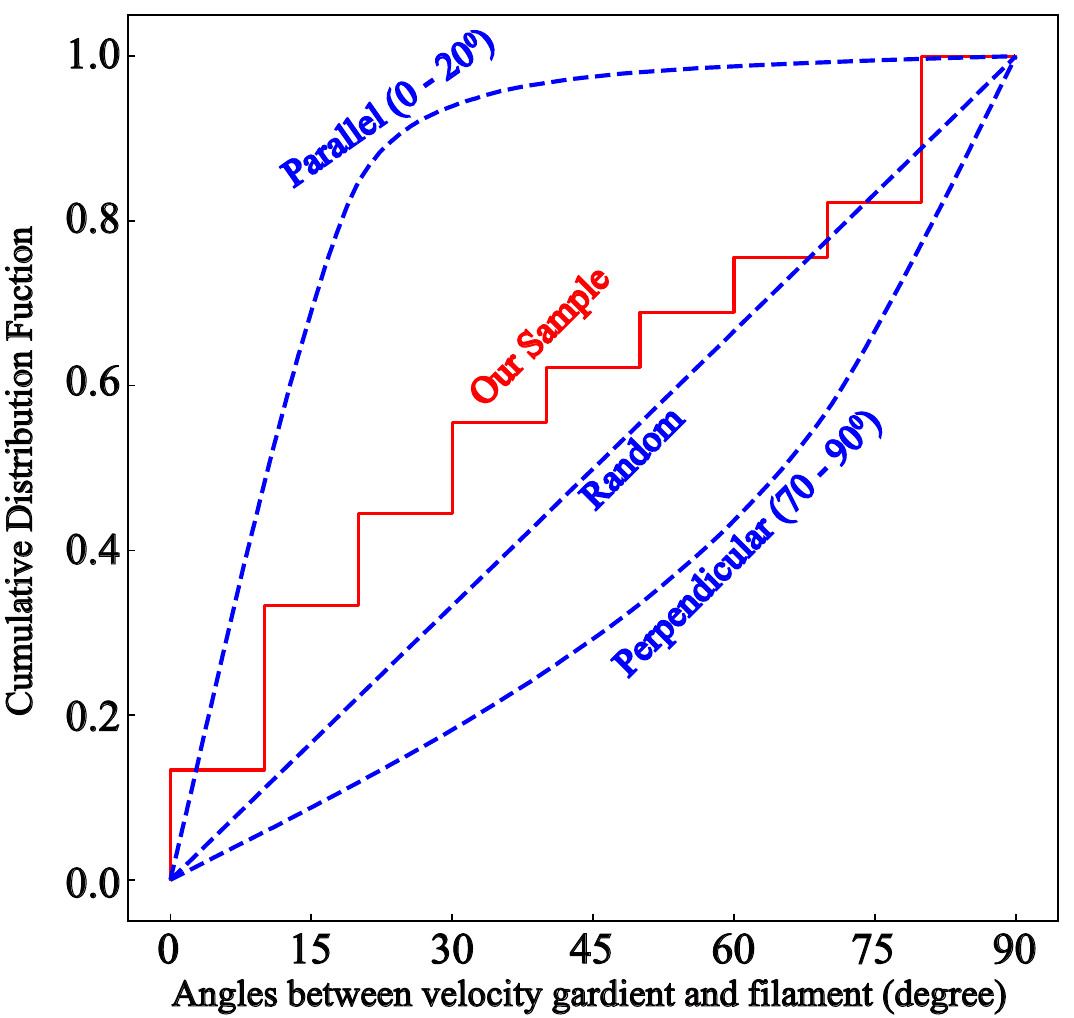}
	\caption{The cumulative distribution function of 
		$\vert \theta_{f} - \theta_{\cal{}G} \vert$ 
		and the projected $\theta_{3D}$. 
		The red step line is the 
		$\vert \theta_{f} - \theta_{\cal{}G} \vert$ of our sample. 
		The three blue dashed lines are results
		from Monte Carlo simulations, which were used to illustrate what an actual aligned sample might look like.  \label{fig:ks}}
\end{figure}

To investigate the distribution of 
$\vert{}\theta_{f} - \theta_{\cal{}G}{}\vert$, Monte Carlo simulations in three-dimensional (3D) space were conducted, following  the approach described by~\cite{Stephens:17}. In this method, two random unit vectors are generated within a unit sphere in 3D, 
and the angle ($\theta_{3D}$) between these two vectors is measured. 
A total of 10$^{6}$ pairs of unit vectors are generated 
to produce 10$^{6}$ angles of $\theta_{3D}$,
constrained to a range of 0$^{\circ}$ -- 90$^{\circ}$. 
If $\theta_{3D}$ exceeds 90$^{\circ}$, the values of 180$^{\circ}$ - $\theta_{3D}$ values are adopted (Equation~\ref{equ6}). 
Angles within 0$^{\circ}$ -- 20$^{\circ}$ are defined as parallel, 20$^{\circ}$ -- 70$^{\circ}$ as random, 
and 70$^{\circ}$ -- 90$^{\circ}$ as perpendicular. 
The angles of $\theta_{3D}$ are then projected onto a two-dimensional (2D) space.
Figure~\ref{fig:ks} plots the cumulative distribution function of 
our $\vert{}\theta_{f} - \theta_{\cal{}G}{}\vert$ 
and the projected $\theta_{3D}$ (the three blue dashed lines), which were  applied to illustrate what our sample might look like. The consistency between the two distributions is assessed using the p-values of the Anderson–Darling (AD) test. A p-value close to 1 suggests a likely consistency between the distributions, whereas a p-value close to 0 indicates their inconsistency. For our dataset, the AD test yields a p-value of 0.98, indicating that the observed distribution of angles bears more resemblance to a random distribution. This finding suggests a non-correlation between the gradient direction of a clump and the orientation of its natal filament. Furthermore, it indicates that the rotation axis of a clump does not depend on the orientation of its natal large-scale filament. 

\section{Discussion} \label{sec:dis}  
\subsection{$J/M$ and $\beta$ in This Study and Previous Studies} \label{sec:angular} 
The derived $J/M$ and $\beta$ in this work were compared with those form earlier reports, which included clumps at various evolutionary stages~\citep[e.g.][]{Phillips:99}, ranging from starless and pre-stellar cores in dark clouds~\citep{Goodman:93,Caselli:02,Xu:20b} to dense cores/clumps in high-mass clouds~\citep{Pirogov:03,Li:12,Tatematsu:16,Xu:20b}. Notably, the $J/M$ and $\beta$ values for 30 well-resolved cores reported by~\citet{Xu:20a} were obtained at the Jeans scale the Orion Molecular Cloud (OMC) 2/3 using high-resolution ALMA (Atacama Large Millimeter/submillimeter Array) N$_{2}$H$^{+}$ images. Additionally,~\citet{Xu:20a} reported a random distribution for $\vert{}\theta_{f} - \theta_{\cal{}G}{}\vert$, consistent with the findings of this study. When considering the complete sample (the values of $J/M$ and $\beta$ from this study and previous studies), $J/M$ and $\beta$ were fitted as functions of clump size $R$, as presented in Figure~\ref{fig:moment}(c). A power-law relation between $J/M$ and $R$ persists across all measurements, yielding a best-fit slope of 1.58, which closely aligns with the value of 1.5 found in this study. The values of $J/M$ for our clumps, represented by green circles in Figure~\ref{fig:moment}(c), are lower than those reported in other studies. This is mainly due to a consistently higher density of our HiGAL clumps , which are distributed along condensed filaments. Other clumps shown in Figure~\ref{fig:moment}(c), even those from high-mass star-formation regions, are often characterized by different observational methods, primarily using CO tracing lower densities. Although $\beta$ shows significant scatter, its small value suggests that rotational energy constitutes only a minor fraction of the gravitational energy, indicating that observed rotation cannot prevent the gravitational collapse of clumps in molecular clouds.

\subsection{The Distribution of $\vert{}\theta_{f} - \theta_{\cal{}G}{}\vert$} \label{sec:angle} 
\begin{figure}
	\centering
	\includegraphics[width=0.48\textwidth]{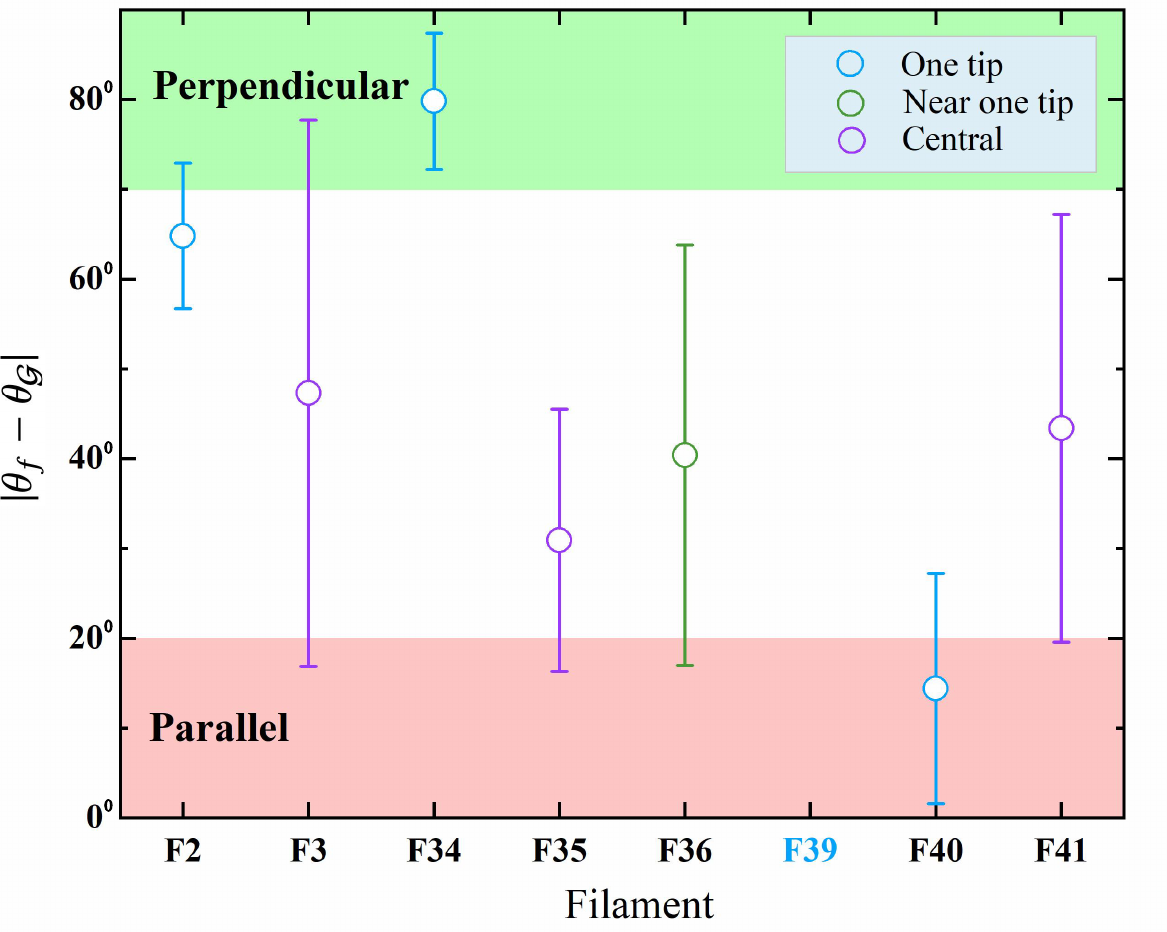}
	\caption{$\vert \theta_{f} - \theta_{\cal{}G} \vert$ for the most massive clump in each target filament. According to the definition of Monte Carlo simulation, the green area corresponds to the perpendicular direction (70$^{\circ}$ $\sim$ 90$^{\circ}$), whereas the light pink area corresponds to the parallel direction, i.e. 0$^{\circ}$ $\sim$ 20$^{\circ}$. For the F39 filament, the most massive clump is at the central, but has no reliably fitted $\theta_{\cal{}G}$. \label{fig:clump-angle}}
\end{figure}

Modifications in the alignment between a clump and its natal filament can reveal their dynamical states, such as collapse, and the manner of gas transfer within filaments. Various angles (e.g., angle among outflow axes, velocity gradients, magnetic fields, disk axes, and filament orientation) between clump and its natal filament have been statistically analyzed to explore any existing trends of alignment. For example,~\citet{Planck:16} and~\citet{Soler:19} studied the relative angles between magnetic fields at local clumps and filament orientations. They found that magnetic fields predominantly align parallel to low column density filaments, whereas they tend to be perpendicular to high column density filaments. At smaller scales,~\cite{Zhang:14} found that the magnetic fields at dense core scales align either parallel or perpendicular to parsec-scale magnetic fields. Furthermore, such bimodal distribution in the angles between the momentum (outflow axes) of clumps  and filament orientation has also been detected in previous works~\citep{Anathpindika:08,Wang:11,Kong:19,Anathpindika:22}. Conversely, a random distribution of the relative angles between momentum and filament orientation/magnetic field has been reported in several recent studies~\citep{Stephens:17,Punanova:18,Xu:20a,Baug:20}. Our finds are consistent with this random distribution, suggesting that the relative orientation between the rotational axis of clumps and filaments might not be as deterministic as previously thought. For the special location, whether located centrally or at one tip, the most massive clump (black solid circles in Figure~\ref{fig:direction}) in each filament also has no obvious preference orientation, 
shown as Figure~\ref{fig:clump-angle}.    

Compared to several previous studies~\citep[e.g.][]{Arzoumanian:18,Zhang:20,Guo:22,Liu:23,Dewangan:24}, our sample is robust because the orientation of the filaments considered here are closely related with the embedded clumps, characterized by the positions of the two tip clumps (see Section~\ref{sec:fila}). In these works, the identification of filaments was carried out employing 2D identification schemes like FILFINDER~\citep{Koch:15}, DisPerSE~\citep{Sousbie:11}, and SExtractor~\citep{Bertin:96}, which often have no direct relation with the embedded cores/clumps. Our experience suggests that these technical differences are unlikely to explain the observed diversity in the relative orientations between the rotational axes of clumps and filaments. A comprehensive study using high spectral and spatial resolution images of a larger sample is required to understand this inconsistency better.

\section{Summary} \label{sec:sum}
The Milky Way atlas for linear filaments~\citep{Wang:24} was integrated with data from the SEDIGISM survey to study the
relative orientations between the rotational axes of clumps and filament orientations.  
The $^{13}$CO (2-1) images from SEDIGISM were used to trace the rotational axes of massive molecular clumps along the filaments. 
The orientation of the filament was characterized by the positions of the clumps at two tips, measured clockwise from the East. Our key finds are summarized as follows:

1. Velocity Gradients and Specific Angular Momentum (J/M): Velocity gradients ranging form 7.0~$\times$~10$^{-2}$~km~s$^{-1}$pc$^{-1}$ to 4.8~km~s$^{-1}$pc$^{-1}$ were derived, corresponding to $J/M$ values between 7.8~$\times$~10$^{-4}$~pc\,kms$^{-1}$ and 7.2~$\times$~10$^{-2}$~pc\,kms$^{-1}$. A monotonic increase in $J/M$ as a function of clump size ($R$) was observed, following a power-law relation $J/M~\propto~R^{1.5\pm0.2}$, similar to the value reported by~\citet{Goodman:93}.

2.Ratio of Rotational to Gravitational Energy ($\beta$): The ratio, $\beta$, was found to range from 1.1~$\times$~10$^{-5}$ to 0.1. Such small values of $\beta$ indicate that rotation alone does not provide sufficient support against the gravitational collapse. 

3.Alignment between Filament Orientation and Velocity Gradient: The angle ($\vert{}\theta_{f} - \theta_{\cal{}G}{}\vert$) between the orientation of the filaments ($\theta_{f}$) and the direction of velocity gradient of clumps ($\theta_{\cal{}G}$) appears to be random, as evidenced by comparisons with the distributions generated via Monte Carlo simulations. Additionally, it was observed that the most massive clump in each filament, whether located centrally or at one tip, exhibited no preferred orientation.

These conclusions enhance our understanding of the dynamics and structural alignments within molecular clouds, suggesting complex interactions rather than deterministic behavior in the formation and evolution of these structures.

\section*{Acknowledgements}

This work is supported by the 
National Natural Science Foundation of China (grant No.\ 11988101, 12103010, U1931104, 12041305 and 12033005), 
Fundamental Research Funds for the Central Universities (Grant No. 2024CDJGF-025), 
Chongqing Municipal Natural Science Foundation General Program (Grant No. cstc2021jcyj-msxmX0867), 
Chongqing Talents: Exceptional Young Talents Project (Grant No. cstc2021ycjh-bgzxm0027),
the National Key R\&D Program of China (2022YFA1603100), 
the China Manned Space Project (CMS-CSST-2021-A09, CMS-CSST-2021-B06), 
the Tianchi Talent Program of Xinjiang Uygur Autonomous Region, 
and the China-Chile Joint Research Fund (CCJRF No. 2211). 
CCJRF is provided by Chinese Academy of Sciences South America Center for Astronomy (CASSACA) 
and established by National Astronomical Observatories, 
Chinese Academy of Sciences (NAOC) 
and Chilean Astronomy Society (SOCHIAS) to support China-Chile collaborations in astronomy. 
DL is a New Cornerstone Investigator.

\section*{Data Availability}
This publication is based on data acquired with the Atacama Pathfinder Experiment (APEX) 
under programmes 092.F-9315 and 193.C-0584. 
APEX is a collaboration among the Max-Planck-Institut fur Radioastronomie, 
the European Southern Observatory, and the Onsala Space Observatory. 
The processed data products are available from the SEDIGISM survey database located at 
https://sedigism.mpifr-bonn.mpg.de/index.html, 
which was constructed by James Urquhart and hosted by the Max Planck Institute for Radio Astronomy.



\bibliographystyle{mnras}
\bibliography{Fila-MN_final} 






\bsp	
\label{lastpage}
\end{document}